\documentclass[preprint, 12pt]{elsarticle}
\usepackage{amssymb}

\usepackage{graphicx}
\usepackage{amsmath}
\usepackage{amsfonts}
\usepackage{physics}
\usepackage{bm}
\usepackage{subcaption}
\usepackage{makecell}
\usepackage{ifthen}
\usepackage{nomencl}
\usepackage{multirow}
\usepackage[commentmarkup=uwave, final]{changes}
\definechangesauthor[color=blue]{sw}

\makenomenclature

\newcommand{\unn}[1][]{%
	\ifthenelse{\equal{#1}{}}{\bm{\hat{u}}(\bm{\theta}, \bm{x}, t)}{\bm{\hat{u}}(\bm{\theta}, \bm{x}_{#1}, t_{#1})}
}
\newcommand{\unnInit}[1][]{%
	\ifthenelse{\equal{#1}{}}{\bm{\hat{u}}(\bm{\theta}, \bm{x}, 0)}{\bm{\hat{u}}(\bm{\theta}, \bm{x}_{#1},0)}%
}
\newcommand{\unnStat}[1][]{%
	\ifthenelse{\equal{#1}{}}{\bm{\hat{u}}(\bm{\theta}, \bm{x})}{\bm{\hat{u}}(\bm{\theta}, \bm{x}_{#1})}%
}
\newcommand{\unnPar}[1][]{%
	\ifthenelse{\equal{#1}{}}{\bm{\hat{u}}(\bm{\theta}, \bm{x}, t)}{\bm{\hat{u}}(\bm{\theta}, \bm{x}, t, {#1})}%
}
\newcommand{\unnStatPar}[1][]{%
	\ifthenelse{\equal{#1}{}}{\bm{\hat{u}}(\bm{\theta}, \bm{x})}{\bm{\hat{u}}(\bm{\theta}, \bm{x}, {#1})}%
}

\newcommand{\smallminus}{\scalebox{0.75}[1.0]{\( - \)}}
\renewcommand{\d}[1]{\;\;\mathrm{d}{#1}}
\renewcommand{\dd}[2]{\;\;\mathrm{d}{#1}\;\mathrm{d}{#2}}

\let\today\relax
\makeatletter
\def\ps@pprintTitle{%
	\let\@oddhead\@empty
	\let\@evenhead\@empty
	\def\@oddfoot{\footnotesize\itshape
		{Accepted Manuscript in Computers \& Fluids} \hfill\today}%
	\let\@evenfoot\@oddfoot
}
\makeatother

\begin{document}
\begin{frontmatter}

\title{Physics-Informed Neural Networks for Parametric Compressible Euler Equations}

\author[inst1]{Simon Wassing\corref{cor1}}

\affiliation[inst1]{organization={German Aerospace Center - Institute for Aerodynamics and Flow Technology \\
                                  Center for Computer Applications in Aerospace Science and Engineering},
                    addressline={Lilienthalplatz~7}, 
                    city={Braunschweig},
                    postcode={38108}, 
                country={Germany}}

\author[inst1]{Stefan Langer}
\author[inst1]{Philipp Bekemeyer}
\cortext[cor1]{corresponding author}
\begin{abstract}
The numerical approximation of solutions to the compressible Euler and Navier-Stokes equations is a crucial but
challenging task with relevance in various fields of science and engineering. Recently, methods from deep learning 
have been successfully employed for solving partial differential equations by incorporating the equations into a loss 
function that is minimized during the training of a neural network. This approach yields a so-called physics-informed 
neural network. It is not based upon classical discretizations, such as finite-volume or finite-element schemes, and
can even address parametric problems in a straightforward manner. This has raised the question, whether physics-informed 
neural networks may be a viable alternative to conventional methods for computational fluid dynamics. In this article 
we introduce an adaptive artificial viscosity reduction procedure for physics-informed neural networks enabling
approximate parametric solutions for forward problems governed by the stationary two-dimensional
Euler equations in sub- and supersonic conditions. To the best of our knowledge, this is the first time that the concept
of artificial viscosity in physics-informed neural networks is successfully applied to a complex system of conservation
laws in more than one dimension. Moreover, we highlight the unique ability of this method to solve forward problems in a 
continuous parameter space. The presented methodology takes the next step of bringing physics-informed neural networks
closer towards realistic compressible flow applications.  
\end{abstract}
\end{frontmatter}
\cleardoublepage
\nomenclature{\(c\)}{differential operator}
\printnomenclature
\section*{Nomenclature}
\subsection*{List of Symbols}
\begin{tabular}{cl}
	$\bm{u}$ & general solution to boundary value problem\\
	$\bm{\hat{u}}$ & neural network output vector\\
	$\Omega$ & spatial domain \\
	$\mathcal{D}$ & general differential operator \\
	$\mathcal{B}$ & boundary condition\\
	$\mathcal{I}$ & initial condition\\
	$\bm{x}$ & vector of euclidean coordinates\\
	$t$ & time\\
	$T$ & upper limit of time\\
	$\rho$ & density\\
	$x$ & first euclidean coordinate\\
	$y$ & second euclidean coordinate\\
	$u$ & velocity in x-direction\\
	$v$ & velocity in y-direction\\
	$E$ & total specific energy\\
	$H$ & enthalpy\\
	$p$ & pressure\\
	$\bm{q}$ & velocity vector\\
	$\bm{W}$ & vector of conserved variables\\
	$\bm{W}_\infty$ & vector of conserved variables at far-field\\
	$\bm{F}_x$ & flux vector in x direction\\
	$\bm{F}_y$ & flux vector in y direction\\
	$\kappa$ & ratio of specific heats\\
	$\tau$ & general parameter of PDE or boundary conditions\\
	$\tau_\mathrm{min}, \tau_\mathrm{max}$ & lower and upper bound of $\tau$\\
	$N$ & number of points for residual evaluation\\
	$N_\infty$ & number of points for Dirichlet/far-field boundary condition\\
	$N_\mathrm{ob}$ & number of points on obstacle surface for wall boundary condition\\
	$\mathcal{L}$ & total loss functional\\
	$\mathcal{L_\mathrm{Res}}$ & residual loss term\\
	$\mathcal{L_\mathrm{I}}$ & initial loss term\\
	$\mathcal{L_\mathrm{B}}$ & boundary loss term\\
	$\mathcal{L_\mathrm{\nu}}$ & viscosity penalty loss term\\
\end{tabular}
\newpage
\begin{tabular}{cl}	
	$M_\infty$ & Mach number at far field\\
	$M$ & local Mach number\\
	$\eta$ & artificial viscosity\\
	$\nu$ & artificial viscosity factor\\
	$\tilde{\nu}$ & prescribed artificial viscosity factor\\
	$M_\mathrm{red}$ & number of reduction epochs\\
	$k$ & order of reduction function\\
	$\alpha$ & weighting factor for initial condition loss\\
	$\beta$ & weighting factor for boundary condition loss\\
	$\gamma$ & weighting factor for viscosity penalty loss\\
	$a$ & semi-major axis of ellipse\\
	$b$ & semi-minor axis of ellipse\\
	$e$ & eccentricity of ellipse\\
	$lr$ & learning rate\\
	$\theta$ & deflection angle for oblique shock\\
	$\delta$ & angle between shock and wall\\
	$C_p$ & coefficient of pressure\\
	$r$ & radius of cylinder\\
	$N_\mathrm{batch}$ & number of point per mini-batch\\
	$\omega^k$ & trainable parameter in adaptive activation layer $k$\\
\end{tabular}	
\vspace{1.5cm}
\subsection*{Abbreviations}
\begin{tabular}{cl}
	PINN & physics-informed neural network\\
	LBFGS & Limited-memory Broyden-Fletcher-Goldfarb-Shanno algorithm\\
	ADAM & adaptive momentum estimation optimization algorithm\\
	PDE & partial differential equation\\
	AD & automatic differentiation\\
	CFD & computation fluid dynamics\\ 
\end{tabular}

\clearpage

\section{Introduction}
The motion of compressible fluids gives rise to nonlinear partial differential equations (PDEs) such as the Euler and 
Navier-Stokes equations. The numerical solution of these equations is for example essential for the development of 
future aircraft configurations. State of the art solvers typically rely on finite volume or finite element algorithms 
\cite{Blazek.2015, Anderson.1994, Langer.2014, Kroll.2016}. 
These algorithms have been developed and refined for decades and they are in regular use for industrial problems such 
as aircraft design. These classical algorithms discretize the domain and thus require the construction of carefully 
crafted computational grids, so-called meshes. Superficially speaking, a finer mesh will result in a more accurate 
result but will also increase the computational cost needed.  Despite a non-negligible progress in efficiency in 
recent years, long term prospects for further acceleration of these codes seem limited. Especially, transferring 
these methods to potentially advantageous hardware like graphic processing units has shown to be challenging. Besides, 
classical algorithms solve a problem for one specific instance of boundary conditions and/or initial conditions. 
Therefore, for certain tasks, such as design optimization, multiple evaluations of the solver are necessary. 
The cumulative computational effort of multiple solver evaluations can be significant and is therefore a possibly 
limiting factor for the usage for commercially relevant problems.\par
Our interest is to investigate numerical methods which have the potential to be applied on future hardware promising 
an acceleration of orders of magnitude. In particular, some initial works indicate that one may be able to transfer 
deep learning based approaches to quantum computers \cite{Kyriienko.2021}, even though the potential for acceleration 
on this hardware is currently still unclear. In addition, to successfully implement such an approach, it is necessary 
to explore the methodological basis that is required for successful algorithmic implementation. With the rising
popularity of machine learning methods and especially deep neural networks, alternative approaches for the solution 
of differential equations, based on methods from this rapidly advancing field have become a popular alternative to 
classical solvers\par
One of these techniques are Physics-Informed Neural Networks (PINNs). The fundamental idea of PINNs is to use a neural 
network as a parametric ansatz function for the approximation of the PDE solution and to optimize the networks 
parameters by minimizing a loss functional which directly incorporates the differential equations as well as the 
initial and boundary conditions. During the training of the network, the loss functional is evaluated at 
random points inside of the domain. PINNs are fundamentally different to classical algorithms because they do not 
rely on a spatial or temporal discretization in a classical sense. Instead a continuous parametric ansatz-function 
(a neural network) is evaluated and optimized at (oftentimes randomly distributed) points in the domain. Hence, the 
task of mesh-generation is replaced by the task to find a point distribution in space and time and an appropriate 
network dimension (i.e. number and width of layers) which, when combined, result in a fast convergence and 
satisfactory final accuracy.\par
Compared to classical methods, PINNs can directly tackle parametric problems. A single network can be trained in a 
continuous parameter space and yield approximate solutions for a whole range of parameter combinations of interest. 
This ability may have implications for aforementioned use cases like design optimization. \par
Similar approaches for approximating PDE solutions via neural networks have been proposed decades ago in works of 
Lagaris~et~al.~\cite{Lagaris.1998} and Dissanayake~et~al.~\cite{Dissanayake.1994}. Even though the idea seems 
natural due to the ability of universal function approximation of neural networks~\cite{Hornik.1989}, the method has 
only recently gained popularity. Nowadays, the networks can be trained more efficiently due to the availability of 
high performance graphics cards. Moreover, the implementation of such algorithms has become straightforward with software 
libraries for deep learning, such as Tensorflow \cite{MartinAbadi.2015} and PyTorch \cite{Paszke.03.12.2019}.  
The term PINN has been introduced by Raissi~et~al.~\cite{Raissi.2019} who demonstrated the use of this approach on a 
number of nonlinear PDEs for forward and inverse problems. Subsequently, PINNs have shown to be applicable for solving
stochastic PDEs~\cite{Zhang.2020}, inverse problems~\cite{Raissi.2019b} and parametric problems~\cite{Hennigh.2021}.
Various alterations of the vanilla PINN formulation have been proposed and improvements for the implementation of 
boundary conditions~\cite{Wang.2022, Sun.2020, Yang.2022}, training facilitation~\cite{Maddu.2022, Wang.2021} and 
training point selection~\cite{Nabian.2021, Lu.2021} have been developed.
	\added{For high frequency problems and large domains, the limited expressibility and the frequency bias of
	neural networks may limit the applicability of PINNs. Therefore, different domain decomposition approaches have been
	proposed, which divide the domain into smaller subdomains and use separate networks to approximate the solution in
	each subdomain. In particular Jagtap et al. have proposed the conservative  PINN~\cite{Jagtap.2020} and extended 
	PINN~\cite{Jagtap.2020d} approaches which use disjoint subdomains and predefined interface conditions to enforce
	continuities at the common interfaces of subdomains. Each network for the respective subdomain is 
	trained with a separate loss function and the training procedures can be parallelized efficiently~\cite{Shukla.2021}.
	On the other hand, soft domain composition approaches such as augmented PINN~\cite{Hu.2023} and finite basis PINNs~\cite{Moseley.2023} use 
	overlapping subdomains and smooth gating or blending functions which locally determine the contribution of each 
	network to a particular point in the domain. 
	The final prediction is the sum of all sub-network predictions, weighted with the window/gating functions.
	Augmented PINNs use additional trainable gating networks while finite basis PINNs use fixed analytical window functions.}
Among many other domains, the flow simulation community has readily adapted PINNs for various cases such as blood 
flow~\cite{Sun.2020, Arzani.2021}, turbulent convection~\cite{Lucor.05.03.2021} and aerodynamics of 
airfoils~\cite{Eivazi.2022}. An important property of PINNs is the fact they allow for a straightforward integration 
of additional available data from various sources such as experiments or higher fidelity simulations into the 
training process. Raissi~et~al.~\cite{Raissi.2020} demonstrate how noisy concentration data of a passive flow agent 
can compensate for incomplete boundary conditions. Even for compressible flows, there has been some success to employ 
PINNs for simple forward and especially inverse problems which exhibit discontinuities 
\cite{Mao.2020, Jagtap.2020, Jagtap.2022}. 
Again, the introduction of additional solution data into the loss is used to compensate for incomplete boundary 
conditions. For a more extensive overview on the usage of PINNs for fluid dynamics the interested reader is referred 
to~\cite{Cai.2021}.\par
Fuks~and~Tchelepi~\cite{Fuks.2020} have identified the need for additional dissipation when solving one dimensional 
hyperbolic conservation laws with PINNs once shocks are present. They pointed out the similarity to classical methods, 
which use artificial dissipation to approximately solve conservation laws. 
	\added{Recently, Coutinho~et~al.~\cite{Coutinho.2023}
	have proposed multiple methods to locally or globally choose artificial viscosity values. They demonstrate 
	the efficacy of their method on one dimensional transient PDEs.}
Here we take this idea to more complex problems by solving the stationary compressible Euler equations in two spatial
dimensions. The compressible Euler equations are a system of conservation laws with four dependent variables (typically 
density, the velocity components and the energy) that describe the behavior of inviscid compressible fluids. 
We also observe that additional dissipative terms are able to facilitate convergence which is the case for supersonic
but also for subsonic problems without shocks as shown in \ref{sec:Apx_non_parametric_results}. To avoid highly dissipative
solutions, we introduce two novel ideas. On the one hand we predict the locally necessary dissipation strength by letting 
the network predict the local viscosity alongside the primitive variables. On the other hand, we use a penalty loss term to 
control dissipation levels during the training. By initially training with high viscosity and then reducing the 
dissipation later, we obtain non dissipative solutions while facilitating convergence during training. 
We do, however, not only solve these equations for one instance of boundary conditions, but rather use the ability of 
PINNs to approximate parametric solutions, essentially extending the dimension of the problems by one or two parameter 
dimensions.  
We choose to restrict this analysis to classical forward problems, meaning that no additional data is incorporated 
into the loss function, besides the information that is available in the form of fully provided boundary conditions. 
So far, the investigation PINNs for complex problems governed by the Euler equations has oftentimes been restricted to 
inverse problems where some form of data of the solution is already provided \cite{Mao.2020, Jagtap.2022}. 
While the ability of incorporating this data is clearly an advantage and possibly one of the most important 
use cases of PINNs, we believe that a solid understanding of the forward problem is necessary for reliably tackling
more complex problems (e.g. higher dimensional).\par
In Sec.~\ref{sec:methods} we give a general introduction to the standard PINN 
approach for solving (parametric) initial and boundary value problems. In Sec.~\ref{sec:compressibleFlowsWithPINNs} 
we discuss the application of the approach to the compressible Euler equations and explain how the incorporation of
artificial dissipation during the training can facilitate convergence when looking at the aerodynamic problem of 
calculating the flow around solid obstacles.\par
In Sec.~\ref{sec:parametric_cylinder} we solve the subsonic flow around an ellipse with a parametric boundary shape and
and variable Mach numbers. In Sec.~\ref{sec:parametric_oblique} we solve the supersonic oblique shock problem with
variable Mach number. Consider also Sec.~\ref{sec:Apx_non_parametric_results}, where non-parametric solutions to the
problems are shown and how PINNs struggle to obtain accurate predictions without artificial viscosity. 
Finally, in Sec.~\ref{sec:conclusion} the presented results are evaluated in a more general context and future implications
as well as newly arising questions are discussed.
The novelties of the paper are twofold. To the best of our knowledge for the first time, we solve parametric problems, 
governed by the two dimensional compressible Euler equations with PINNs. In addition we introduce a novel adaptive viscosity
training procedure which improves prediction accuracy of PINNs on supersonic problems significantly, compared to previously 
published results. 
\added{
	Compared to previous work in \cite{Jagtap.2022} we focus exclusively on forward problems. In \ref{sec:Apx_non_parametric_results}, we show that these require additional measures such as artificial viscosity
	to reliably obtain physically reasonable solutions. This has also been shown for more simple conservation laws 
	in Fuks~and~Tchelepi~\cite{Fuks.2020} and Coutinho et al.~\cite{Coutinho.2023} confirm the efficacy of artificial 
	viscosity on one-variable conservation laws. 
	Compared to the work on forward problems by Mao et al. \cite{Mao.2020} we focus on parametric and thus more 
	complex forward problems and obtain better resolution of shocks in the parametric formulation, compared to their
	non-parametric result (see Fig.~\ref{fig:shockduringtraining}). For more details, please consider Sec.~\ref{sec:Apx_non_parametric_results}. 
}
Preliminary results of the presented ideas have been presented at the 8th European Congress on Computational Methods
in Applied Sciences and Engineering~\cite{Wassing.5th9thJun2022}.
\begin{figure}
    \centering
    \includegraphics[width=.8\textwidth]{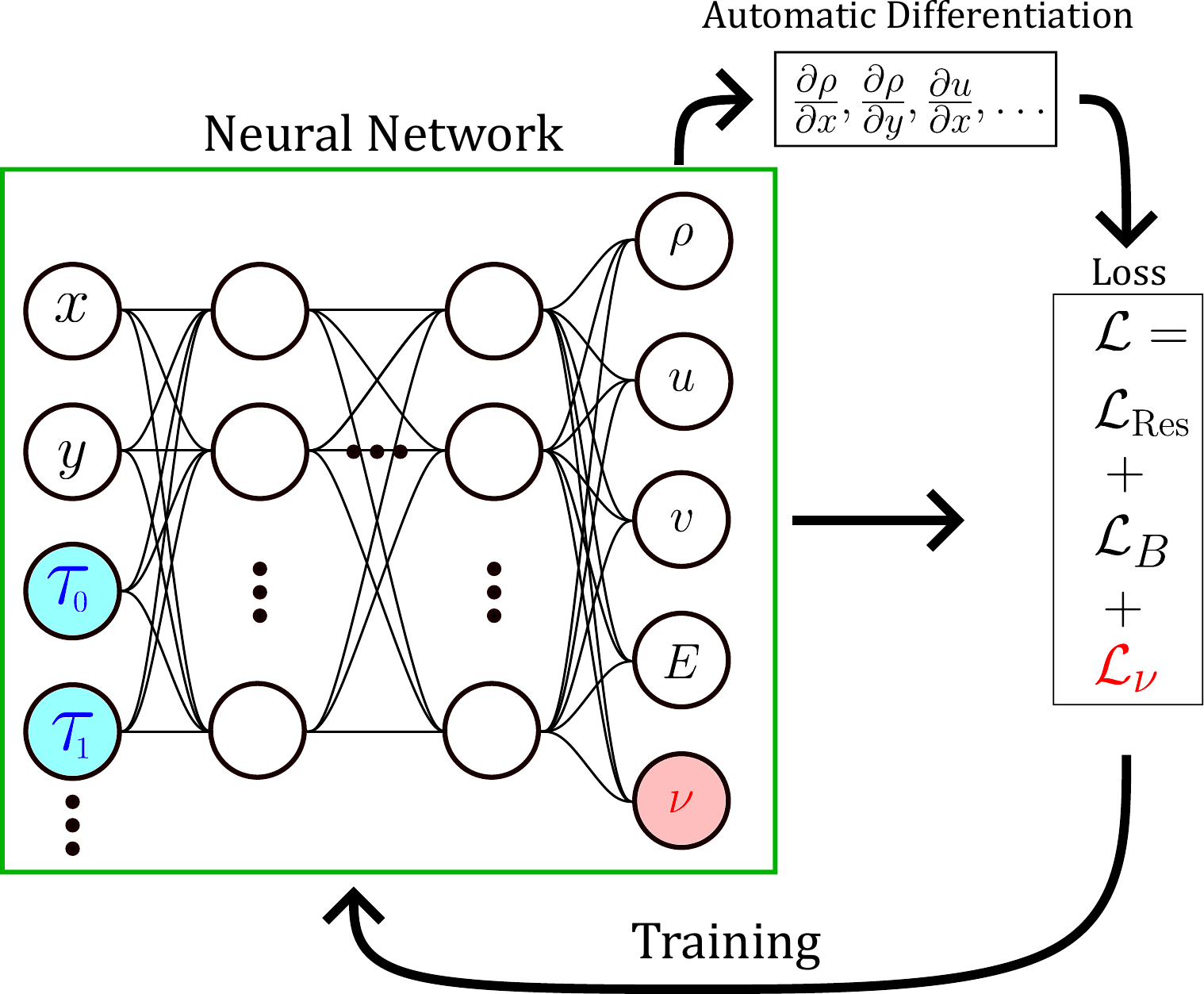}
    \caption{Schematic representation of the physics-informed neural network approach for solving the 2D stationary
    Euler equations. 
    A fully connected neural network predicts the primitive variables $\rho, u , v, E$ and the local viscosity 
    factor $\nu$ at a point $(x, y)$ in the physical domain and for optional parameters of the problem
    ($\tau_0, \tau_1, \dots$). 
    The loss is calculated as the sum of squares of the residual of the PDE, the boundary condition and an additional
    viscosity penalty term. Partial derivatives of the primitive variables with respect to 
    $x$ or $y$ are required for the loss and can be calculated using automatic differentiation.}
    \label{fig:PINN}
\end{figure}

\section{Methods}
\label{sec:methods}
\subsection{Physics-Informed Neural Networks}
Physics-informed neural networks are deep neural networks, which are employed to approximate solutions to differential 
equations. A general initial-boundary value problem for the unknown function $\bm{u}$ on the spatial domain
$\Omega\subset\mathbb{R}^d$ and in the time interval $(0, T)\subset\mathbb{R}$ can be defined as:
\begin{equation}
    \begin{aligned}
        \mathcal{D}(\bm{u}(\bm{x},t), \bm{x},t) &= 0\;\; &  &(\bm{x},t) \in (\Omega \times (0,T))\\
        \mathcal{B}(\bm{u}(\bm{x},t), \bm{x},t) &= 0\;\; &  &(\bm{x},t) \in (\partial\Omega \times (0,T))\\
        \mathcal{I}(\bm{u}(\bm{x},0), \bm{x}) &= 0\;\;   &  &\bm{x}     \in \Omega,
    \end{aligned}
    \label{eq:PDE}
\end{equation}
where $\mathcal{D}$ is a general differential operator and $\mathcal{I}$ and $\mathcal{B}$ are the initial and 
boundary condition, respectively. The operator $\mathcal{D}$ may include multiple nonlinear differential terms of 
different order. A neural network $\unn$ is now used to approximate the unknown solution 
$\unn \approx \bm{u}(\bm{x},t)$. The vector $\bm{\theta}$ includes neural networks parameters (weights and biases) 
which have to be adjusted during an optimization/training process to find an accurate approximation of the solution.
For this, an objective or loss functional $\mathcal{L}$ is defined. If $\unn$ is a solution to Eqs.~\ref{eq:PDE}, 
the left hand side of Eqs.~\eqref{eq:PDE} vanishes on the entire domain $x\in\Omega$ and for all times 
$t\in (0,T)$. Therefore, a simple loss functional is
\begin{eqnarray}
  \mathcal{L}(\bm{\hat{u}}(\bm{\theta})) & = & \int_0^T\int_{\Omega} \mathcal{D}(\unn)^2\dd{\bm{x}}{t}\, +
  \alpha\int_{\Omega}\mathcal{I}(\unnInit, \bm{x})^2 \d{\bm{x}}\,\nonumber\\
  & + &	\beta\int_0^T\int_{\partial\Omega} \mathcal{B}(\unn,\bm{x},t)^2\dd{s(\bm{x})}{t} \label{eq:lossInt}
\end{eqnarray}
which is 0 and thus minimal if $\unn$ is a solution to Eqs.~\ref{eq:PDE}.
The coefficients $\alpha$ and $\beta$ scale the importance of the boundary and initial loss term with respect to the 
residual term. Since the network $\unn$ can only be evaluated at discrete input values, the loss is instead calculated
by taking the sum over a representative point distribution inside of the spatial and temporal domain
\begin{equation}
    \begin{aligned}
        &\mathcal{L}(\unn) = \mathcal{L}_\mathrm{Res} \;+\; \mathcal{L}_\mathrm{I} \;+\; \mathcal{L}_\mathrm{B}&&\\
        &\mathcal{L}_\mathrm{Res}=\frac{1}{N}\sum_{i=1}^{N}\mathcal{D}(\unn[i])^2\; &&\bm{x}_i\in\Omega\;;t_i\in(0,T)\\
        &\mathcal{L}_\mathrm{I}=\alpha\frac{1}{N_\mathrm{I}}\sum_{i=1}^{N_\mathrm{I}}\mathcal{I}(\unnInit[i])^2\; &&\bm{x}_i\in\Omega\\
        &\mathcal{L}_\mathrm{B}=\beta\frac{1}{N_\mathrm{B}}\sum_{i=1}^{N_ \mathrm{B}}\mathcal{B}(\unn[i])^2	\; &&\bm{x}_i\in\partial\Omega\;;t_i\in(0,T)
    \end{aligned}
    \label{eq:lossSum}
\end{equation}
where $N_\mathrm{Res}$, $N_\mathrm{I}$ and $N_\mathrm{B}$ are the number of points that evaluate the residual, the 
initial condition and the boundary condition, respectively. Since for a well defined problem, the bounds of the 
domain are known, the generation of these training points can be achieved with quasi-random low discrepancy sequences
such as Sobol~\cite{Sobol.1967} or Halton~\cite{Halton.1960} or with other methods like Latin Hypercube 
sampling~\cite{McKay.1979} at little additional cost. For a uniform point distribution, the calculation of the loss 
functional can be interpreted as a Monte-Carlo integration of Eq.~\eqref{eq:lossInt} where the normalization by the
volume has been dropped. As highlighted in various publications~\cite{Lu.2021, Nabian.2021, Mao.2020}, a non uniform
distribution of training points or an adaptive training point selection, based on the local residual may accelerate 
training and improve the final accuracy for certain problems. \par 
The partial derivatives in $\mathcal{D}(\unn[i])$ are calculated, using automatic differentiation~(AD). 
The network parameters are then tuned using an iterative optimization algorithm. The most popular algorithms are 
variants of stochastic gradient descend such as Adam~\cite{Kingma.22.12.2014}. During the optimization, the gradient 
of the loss with respect to the network parameters $\nabla_{\bm{\theta}}\mathcal{L}(\unn[i])$ has to be calculated.
This can again be achieved, using automatic differentiation and the backpropagation algorithm~\cite{Rumelhart.1986}.
As usual for neural networks, the optimization problem is generally non-convex and the optimizer can therefore converge
to local minima or regions of vanishing gradients. Furthermore, for partial differential equations such as the 
compressible Euler equations which may not have a unique solution, a minimum of the loss may not always correspond to 
a physically reasonable solution.
\vspace{.5cm}
\subsubsection{Adaptive Activation Functions}
\added{
A general feedforward neural network with $m$ layers and $d_k$ neurons per layer can be described as a composition of
linear functions and nonlinear activation functions $\sigma$:
\begin{equation}
	\begin{aligned}
		\hat{u}(&\bm{x}^0): \mathbb{R}^{d_0} \longrightarrow \mathbb{R}^{d_m}\\
		\hat{u}_j &\equiv x_j^m \hspace{1cm} j=1,2,\dots ,d_m\\
		\bm{x}^0 &= (x^0_1, x^0_2,\dots, x^0_{d_0}) \\
		x_j^{k} &= \sigma^k\left( \sum^{d_{k-1}}_{i=1} w_{i,j}^{k-1}x_{i}^{k-1} - b_j^k\right)\\
		j&=1,2,\ldots, d_k\;\;\; k=1,2,\ldots, m.
	\end{aligned}
	\label{eq:nn}
\end{equation}
A popular choice for the activation function for PINNs is the hyperbolic tangent which has been shown to be a robust 
choice on a variety of problems~\cite{Jagtap.2023} compared to other fixed activation functions:
\begin{equation}
	\begin{aligned}
		\sigma(x)^k&\equiv\tanh(x)=\dfrac{e^x-e^{-x}}{e^x+e^{-x}}
		\\k&=1,2\ldots, m\smallminus 1. 	
	\end{aligned}
\end{equation} 
Note that in the last layer, the identity function $\sigma^m (x) = x$ is used. 
In this work we also make use of the layer-wise locally adaptive hyperbolic tangent activation formulation~\cite{Jagtap.2020b}:
\begin{equation}
	\begin{aligned}
		\sigma_{\mathrm{adapt}}^k(x)&\equiv\sigma(n\omega^k\cdot x)=\tanh(n\omega^k\cdot x)  \\k&=1,2\ldots,  m\smallminus 1,
	\end{aligned}
\end{equation} 
where $n$ is a constant scaling factor and $\omega^k$ is an additional trainable parameter per hidden layer that scales 
the slope of the hyperbolic tangent. Adaptive activation functions have been shown to improve convergence speed 
\cite{Jagtap.2020c, Jagtap.2020b} and PINN accuracies~\cite{Jagtap.2023} on various problems, compared to fixed 
activation functions. The above described version only introduces a single trainable parameter per layer which 
only marginally increases the computational effort compared to fixed activation functions. 
}
\vspace{.5cm}
\subsubsection{Parametric Problems}
One prospect of physics-informed neural networks is the possibility to solve parametric problems. Let $\tau$ be a 
general parameter of the initial and boundary value problem. A PINN approximation $\unnPar[\tau]$ of the unknown 
solution $u(\bm{x}, t, \tau)$ simply receives $\tau$ as an additional input to the neural network alongside $\bm{x}$
and $t$. The parameter adds an additional dimension to the input space and thus the training points. The training 
points are now sampled in a $d+2$-dimensional domain 
$(\bm{x}, t, \tau) \in \Omega\times(0, T)\times(\tau_\mathrm{min}, \tau_\mathrm{max})$. 
The parameter $\tau$ can then be incorporated into the calculation of any of the loss terms in Eq.~\eqref{eq:lossSum}. 
Similarly, this approach can be extended to more than one parameter.
\subsection{Approximation of Stationary Compressible Flows with PINNs}
\label{sec:compressibleFlowsWithPINNs}
The inviscid flow of compressible fluids is governed by the Euler equations which describe the conservation of mass,
momentum and energy in a continuous fluid. The two-dimensional Euler equations in their differential conservative 
form are given by
\begin{equation}
    \begin{aligned}
        \pdv{\bm{W}}{t} + \pdv{\bm{F}_x}{x}+\pdv{\bm{F}_y}{y} &= 0,\\
        \bm{W} = 
        \begin{pmatrix}
            \rho \\
            \rho u \\
            \rho v \\
            \rho E
        \end{pmatrix},\,
        \bm{F}_x = 
        \begin{pmatrix}
            \rho u\\
            \rho u^2 + p \\
            \rho u v \\
            \rho H u \\
        \end{pmatrix},\, 
        \bm{F}_y &= 
        \begin{pmatrix}
            \rho v \\
            \rho u v \\
            \rho v^2 +p \\
            \rho H v \\
        \end{pmatrix}\\
    \end{aligned}
    \label{eq:euler}
\end{equation}
where $\bm{W}$ is the vector of the conservative variables with the density $\rho$, the local fluid velocity 
$\bm{q} = (u, v)$, and the total specific energy $E$. The total enthalpy $H$ is defined as $H=E+\frac{p}{\rho}$.
This system of partial differential equations can be closed by the equations of state of ideal gases, which yield
$p=\rho (\kappa-1) (E - \rho \bm{q}^2/2)$, with $\kappa$ being the ratio of specific heats ($\kappa = 1.4$ for air).
The Mach number $M=\|\bm{q}\|_2/a$ is the ration between the velocity and the speed of sound 
$a=\sqrt{\kappa p/\rho}$.\par
For aerodynamic problems one is oftentimes interested in the steady-state solution. Therefore, the time derivative in
Eq.~\eqref{eq:PDE} is omitted.\par
When solving this system of equations with a physics-informed neural network, the network outputs can be chosen to 
approximate the primitive variables $\unnStat \approx (\rho(\bm{x}), u(\bm{x}), v(\bm{x}), E(\bm{x}))$. 
The~residual~loss~term~becomes
\begin{equation}
    \begin{aligned}
        \mathcal{L}_\mathrm{Res} = \dfrac{1}{N}\sum_{i=1}^{N} \left(\pdv{\bm{F}_x(\unnStat[i])}{x}+\pdv{\bm{F}_y(\unnStat[i])}{y}\right)^2.
    \end{aligned}
\end{equation}
The partial derivatives are calculated using automatic differentiation.\par
In the following, we focus on the calculation of flows around solid obstacles and alongside walls. In particular, we use 
Dirichlet boundary conditions for the inflow/far-field and wall boundaries for the obstacles. 
For the Dirichlet boundary conditions the conserved variables approach 
$\bm{W}_\infty = (\rho_\infty, \rho_\infty u_\infty, \rho_\infty v_\infty, \rho_\infty E_\infty)$ which is in the 
context of aerodynamics typically determined by the Mach number $M_\infty$. Also, the flow should be tangential to the 
obstacles surface. This results in a boundary loss term of
\begin{equation}
    \begin{aligned}
        \mathcal{L}_\mathrm{B} =\beta \bigg( \dfrac{1}{N_\infty}&\sum_{i=1}^{N_\infty} \left[\bm{W}(\unnStat[i])- \bm{W_\infty}\right]^2 + \\ 
        \dfrac{1}{N_\mathrm{ob}} &\sum_{j=1}^{N_\mathrm{ob}}\left[\bm{q}(\unnStat[j])\cdot \bm{n}_j\right]^2 \bigg)
    \end{aligned}
\label{eq:BoundaryLossEuler}
\end{equation}
with $N_\infty$ points $\bm{x}_i$ on the boundary of the physical domain and $N_\mathrm{ob}$ points $\bm{x}_j$ on
the boundary of the obstacle. The surface normals at positions $\bm{x}_j$ are given by $\bm{n}_j$.
\begin{figure}
    \centering
    \includegraphics[width=\textwidth]{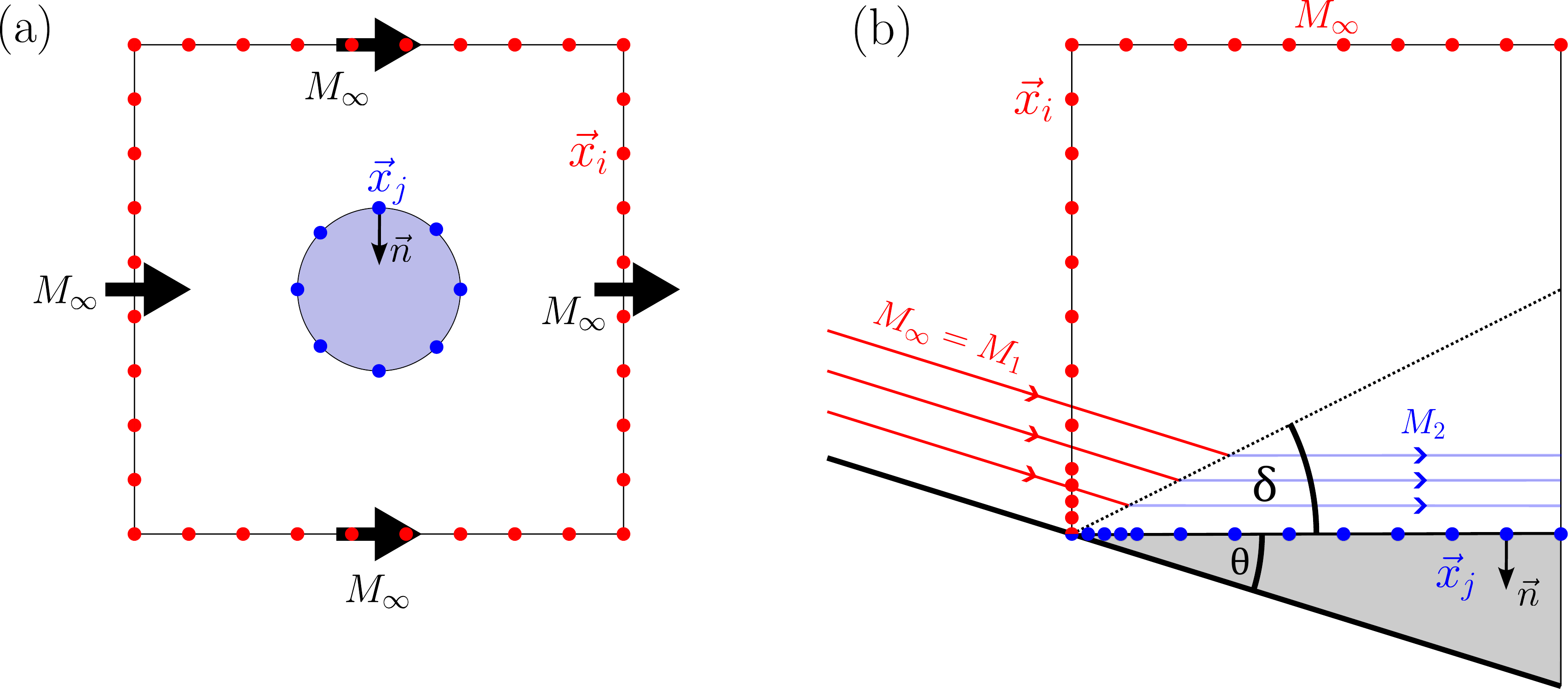}
    \caption{Schematic representation of boundary conditions and boundary training points for the cylinder
    and the oblique shock test case. 
    The Dirichlet boundary conditions for a Mach number of $M_\infty$ are applied at the points $\bm{x}_i$ with 
    $i=1\dots N_\infty$. The wall boundary conditions at the obstacle surfaces are enforced at the points 
    $\bm{x}_j$ with $j=1\dots N_\mathrm{ob}$. Note that for the parametric problems, one additional sampling dimension 
    per parameter is appended to the euclidean spatial coordinates.}
    \label{fig:BCs}
\end{figure}
\subsubsection{Adaptive Artificial Dissipation}
A fundamental challenge for PINNs in the search of solutions to the compressible Euler equations lies in finding a 
physically reasonable minimum of the loss (i.e. an entropy solution). By definition of the loss 
function~\eqref{eq:lossSum}, any weak solution of Eqs.~\eqref{eq:euler} that fulfills a given set of boundary
conditions is a (global) minimum of the loss and thus a possible final state of the network to converge to.
Furthermore, the optimizer can always converge to local minima, which correspond to unphysical solutions.\par
In classical computational methods for the solution of compressible flows, artificial dissipation is introduced in 
the form of upwind schemes or through a combination of central schemes and explicit dissipative terms 
\cite{Blazek.2015}. The dissipation smears out discontinuities and has a stabilizing effect.\par
For PINNs, the necessity of additional dissipative terms for successfully solving scalar conservation laws has also 
been observed by Fuks et al. \cite{Fuks.2020}. More complex systems of conservation laws, such as the compressible
Euler equations require advanced methods for locally determining reasonable dissipation levels while also minimizing
the dissipative effects on the solution. \par
Here we propose a novel training procedure for flexibly and reliably solving the compressible Euler equations which 
uses artificial dissipation. The novelty of this procedure is twofold. Firstly, the PINN locally predicts an
appropriate value of viscosity which is necessary to stabilize the training. Secondly, the strength of dissipation 
is reduced during the training process which minimizes errors that are introduced by the dissipation. \par
The dissipation is introduced as an additional term in Eqs.~\eqref{eq:euler}:
\begin{equation}
	\begin{aligned}
		\pdv{\bm{W}}{t} + \pdv{\bm{F}_x}{x}+\pdv{\bm{F}_y}{y} &= \eta \Delta \bm{W}
	\end{aligned}
	\label{eq:eulerDissipation}
\end{equation}
where $\Delta=(\partial_x^2 + \partial_y^2)$ is the Laplacian and $\eta$ can be interpreted as an artificial viscosity.
From an optimization point of view, the dissipation acts as a regularization of the optimization/training problem. 
Similarly to classical scalar dissipation schemes like the JST-scheme~\cite{JAMESON.1981}, we scale the dissipation 
locally, based on the spectral radius of the flux jacobians
\begin{equation}
	\eta = \nu (a + |\bm{q}|)
	\label{eq:artVisc}
\end{equation}
where $\nu$ is a viscosity factor for which adequate values have to be selected. The linear scaling of the viscosity 
$\eta$ with the wave speed $a+|\bm{q}|$ results in higher viscosity values in regions with higher Mach numbers. 
Furthermore, when considering parametric problems with variable inflow velocities, the viscosity is adjusted according 
to the resulting variable wave speed. 
However, additional local upscaling of the viscosity may be required to deal with certain highly nonlinear or unstable 
regions in the domain, such as shocks. Instead of deriving an analytical sensor function, we leverage the
ability of the neural network to adjust the viscosity. 
To do so, the network predicts the viscosity factor $\nu$ alongside the primitive variables 
$\unnStat \approx (\rho(\bm{x}), u(\bm{x}), v(\bm{x}), E(\bm{x}), \nu (\bm{x}))$. 
To enforce positivity of the viscosity and to allow for a flexible prediction of values close to 0, the exponential 
function is used as an activation for $\nu$ in the last layer of the network. In addition, since we are interested 
in the inviscid solution, an additional loss term $\mathcal{L}_\nu$ is introduced which 
penalized high viscosity values : 
\begin{equation}
	\mathcal{L}_\nu = \gamma|\nu - \tilde{\nu}|.
\end{equation}
The prescribed viscosity value $\tilde{\nu}$ can now be used to control the strength of the dissipative term, while 
still allowing the network to locally choose $\nu(\bm{x})\neq \tilde{\nu}$ when necessary. We use the modulus to 
enforce positivity of the term instead of the square because local outliers should not be penalized.
We want to stress that $\tilde{\nu}$ is not related to a turbulent eddy viscosity for which this symbol is oftentimes 
used in turbulence models.\par 
The addition of dissipation changes the physical problem and the resulting solution will disagree with the fully 
inviscid solution. Essentially, one is no longer solving for an inviscid solution and therefore the fluid is 
decelerated by shear stresses. In classical methods, scalar dissipation schemes employ fourth order differences in 
smooth flow regions which only dampens higher frequency modes of the flow, limiting the effect of the artificial 
viscosity on the solution. Second order differences are only used near discontinuities, where a scheme of first order 
accuracy is required due to Godunov's theorem. An additional fourth order differential term could theoretically be 
added to Eq.~\eqref{eq:eulerDissipation}. This would however introduce fourth order derivatives which would require 
additional expensive evaluations of the computational graph. Instead, we follow a different approach and adjust 
the viscosity during the training process, to limit the dissipative effect on the final solution. Therefore, we 
propose a 3 phase training routine:
\begin{itemize}
	\item[] Phase 1: Train with ADAM at a constant viscosity $\tilde{\nu}=\nu_0$ until no significant changes in the residual loss are observed.
	\item[] Phase 2: Train with ADAM and reduce the prescribed viscosity $\tilde{\nu}$ until $\tilde{\nu}=0$. Continue at $\tilde{\nu}=0$  until no significant changes in the residual loss are observed.
	\item[] Phase 3: Train with LBFGS and $\tilde{\nu}=0$ until convergence.
\end{itemize}
The idea of this procedure is to guide the network towards a physical solution (the entropy solution) during the
initial training phase. Once the network has converged to a state that resembles a physical but viscous solution, the
viscosity can be reduced, since, from this point on, the network should be in a state near the entropy solution. 
In phase 2, $\tilde{\nu}$ is reduced to 0, to decrease the dissipative effect on the prediction. 
However, the network can still predict viscosity factors $\nu(\bm{x})>0$ where necessary, keeping the penalty term in
balance with the residual and other loss terms.
We reduce the prescribed viscosity factor $\tilde{\nu}$ as follows:
\begin{equation}
	\tilde{\nu}_i=
	\begin{cases}
		  \nu_0 \bigg( 1- \bigg(\frac{i}{M_{\mathrm{red}}}\bigg)^k \;\bigg) & \text{if}\;\; 0 \leq i \leq M_\mathrm{red}\\
		  0 & \text{if}\;\; M_\mathrm{red} < i 
	\end{cases}
\end{equation}
Where $i\in \mathbb{N}$ is the epoch counter starting at phase 2 and $M_{\mathrm{red}}$ is the epoch at which 
$\tilde{\nu}=0$ is reached. The exponent $k\in \mathbb{N}^+$ can be used to modify the shape of the reduction curve. 
For $k=1$ the reduction is linear and for $k>1$ it is accelerating.
For the first two phases, Adam~\cite{Kingma.22.12.2014} is used as the optimizer. A final training period with the 
quasi-Newton LBFGS optimizer~\cite{Liu.1989} has shown to be effective for convergence of PINNs in general 
and has been crucial to achieve high levels of accuracy with the proposed training procedure.
A schematic representation of the described PINN approach is shown in Fig.~\ref{fig:PINN} and a schematic view of 
the boundary points for the later discussed test cases is shown in Fig.~\ref{fig:BCs}.\par
\vspace{.5cm}
\section{Parametric Flow around Ellipse}
\label{sec:parametric_cylinder}
As the first parametric problem we consider the flow around an ellipse with a parametric inflow boundary 
condition and a parametric ellipse boundary. We position the two-dimensional ellipsoid at the
center of the domain $\Omega = (-1,1)\times(-1,1)$. The semi-major axis $a\in(0.1, 0.2)$ is variable whereas the semi
minor axis $b=0.1$ is constant. Therefore at a value of $a=0.1$, we have a cylinder of radius 
$r=0.1$ which corresponds to an eccentricity of $e=\sqrt{1-b^2/a^2} = 0$. For the maximal major axis of $a=0.2$ the
eccentricity is $e=3/4$. As a second varying parameter we consider the Mach number $M_\infty\in(0.2, 0.4)$. The 
neural network receives both parameters as additional inputs. The training points are therefore sampled in a 
four-dimensional domain. The upper limit of the Mach number of $M_\infty=0.4$ is slightly below the critical Mach
number of the cylinder, at which the velocity locally exceeds the speed of sound. An initial prescribed artificial 
viscosity factor of $\nu=7.5\cdot10^{-4}$ is used based on the previous investigations. We have observed that this 
value is typically a reasonable value and can be used both for subsonic and supersonic problems. 
Since the actual artificial viscosity in Eq.~\eqref{eq:artVisc} is scaled with the local wave speed, the strength of 
the dissipation naturally increases at higher Mach numbers. \par
We choose a fully connected neural network of constant layer width.
\deleted{
	Adaptive activation functions (see e.g. \cite{Jagtap.2020b}) have been reported to improve prediction accuracy and 
	to speed up convergence in PINNs. For the analyzed problems we could observe improved convergence early during the 
	training, when using layer-wise adaptive activations. However, when using the LBFGS optimizer during the final
	training phase, they have performed inconsistently (i.e. on some runs similarly to $\tanh$ and on others very poorly). 
	Therefore, we continue to use standard hyperbolic tangent activations.}
\added{We investigate the performance of fixed and adaptive hyperbolic tangent activation functions. We observe
    that during the third training phase, when using the LBFGS optimizer, the usage of adaptive
    activation functions leads to highly inconsistent results (i.e. in some runs convergence is similarly to $\tanh$
    and in others no further convergence is possible or the optimizer diverges completely).
    Therefore, we decide to freeze the trainable parameters $\omega^k$ for the final training phase meaning that these 
    parameters are no longer optimized during phase~3. In doing so, we are able to avoid the previously observed inconsistencies.}
\par
The loss weighting factors are set to $\beta=1$ and $\gamma=5$. As shown in various publications, such as
\cite{Jin.2021, Maddu.2022, Wang.2021}, the (dynamic) weighting of loss terms is an effective technique to accelerate
the convergence and improve the accuracy because it can compensate imbalances in the gradients between the different
loss terms. However, we have observed that for certain problems, adaptive loss term weighting may lead to instabilities
during the early stages of the training. For the presented problems we do not see significant imbalances during the
training and are able to achieve satisfying results with a constant weighting factor. Therefore, dynamic
loss term weighting is not considered herein.\par
\begin{figure}
	\centering
	\includegraphics[width=.5\linewidth]{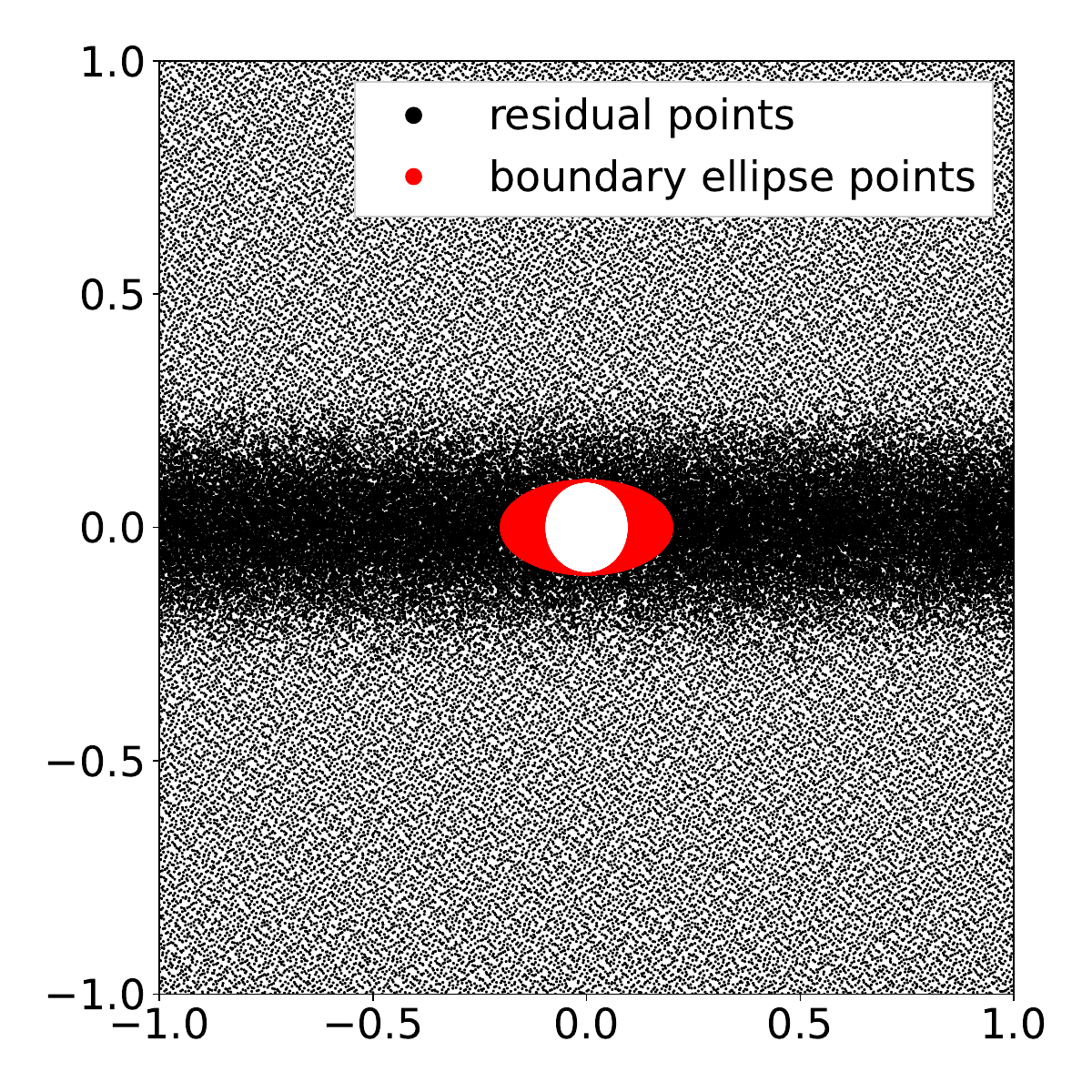}
	\caption{Projection of point distribution of four-dimensional parametric space onto physical domain for subsonic 
		ellipse problem. Note the variation in the axis $a$ in the boundary points.}
	\label{fig:pointDistrParametric}
\end{figure}
\begin{figure}[p]
    \centering
    \includegraphics[width=1\textwidth]{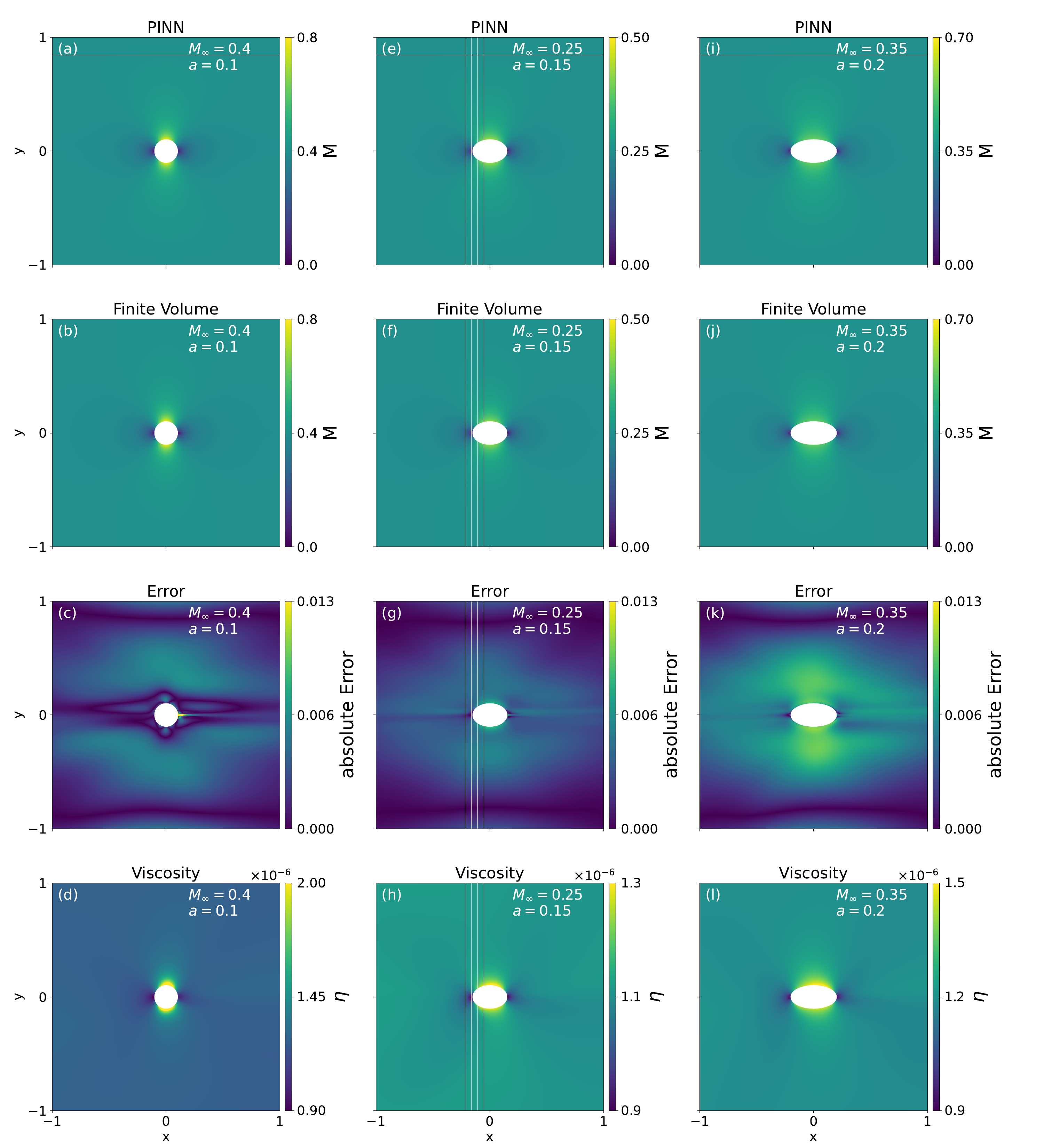}
    \caption{\comment{(updated) }Comparison between parametric PINN solution \added{
    with adaptive activation functions} and a 
    reference finite volume result for different Mach numbers and ellipse eccentricities.
    The absolute errors between the reference and the PINN solution are shown in
    Figs.~(c),~(g) and (k). Figs.~(d),~(h)~and~(l), show the artificial viscosity $\eta$.}
    \label{fig:ParametrizedMaNew}
\end{figure}

A summary of all hyperparameters is shown in Tab.~\ref{tab:training_parameters_cylinder}. 
This includes the utilized optimizer, the learning rate $lr$ the batchsize $N_\mathrm{batch}$ and the prescribed 
viscosity factor $\tilde{\nu}$.
To cover the four-dimensional input space, the number of residual training points is comparatively high ($N=100000$). 
However, since a mini-batch routine is used for the first two training phases, this has typically no negative effect 
on the training speed. Only during the last training phase 3, memory may be a limiting factor because the LBFGS 
optimizer is incompatible with mini-batch training. Therefore, we use a reduced number of training points during the 
last training phase ($N=30000$). 
A non-uniform point distribution is used. Half of the points are
distributed uniformly across the entire physical domain $\Omega= (-1,1)\times(-1,1)$ using the Halton sequence 
\cite{Halton.1960}. For the other half of the points, the y-coordinate is sampled using a normal distribution with a
variance of $\sigma=0.07$ and with a uniform distribution for the x-coordinate. A projection of the resulting point
distribution to the physical domain is shown in Fig.~\ref{fig:pointDistrParametric}.
For both point sets, the same number of points is used to represent the boundary of the physical domain and the 
cylinder ($N=N_\mathrm{\infty}=N_\mathrm{ob}$). No additional data of the solution is incorporated into the loss, 
besides the boundary conditions. 
We are thus approximating the solution of a fully determined (but not over-determined) forward problem.  
The resulting predictions for the velocity field for three 
cases are shown in Fig.~\ref{fig:ParametrizedMaNew} in comparison to reference finite volume simulations. 
For additional information on the calculation of reference finite volume results, see
Sec.~\ref{sec:apx_simulations_and_errors}. One can see that for the cylindrical shape ($a=0.1$) even at $M_\infty=0.4$, 
close to the critical Mach number,  the results are visually indistinguishable to the reference solution. 
The plot of the absolute error reveals that the inaccuracies are fairly uniform. 
For the other two parameter sets with ellipsoidal shapes, a similar quality of the results can be observed.
The bottom row of Fig~\ref{fig:ParametrizedMaNew} depicts the artificial viscosity $\eta$. The viscosity is relatively
uniform (see the scale of the color bars) for all three Mach number. Up and downstream of the ellipse, the viscosity
is however slightly reduced.\par
\begin{figure}
	\centering
	\includegraphics[width=.8\linewidth]{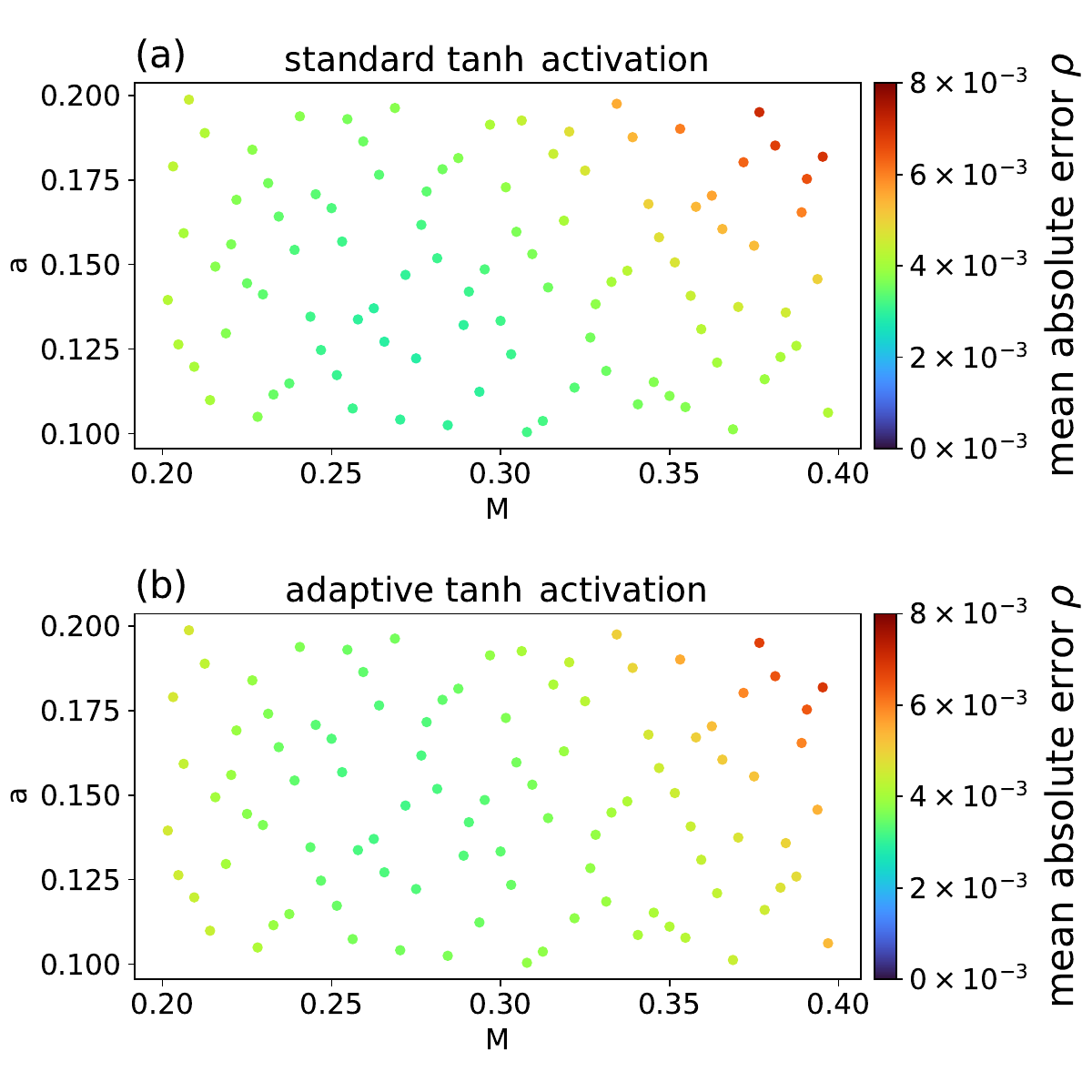}
	\caption{\comment{(updated) }Mean absolute errors on subsonic test case between parametric PINN and reference finite volume result for 
		the density field, for different parameter sets. Fig.~(a) shows the results without and Fig.~(b) with adaptive 
		$\tanh$ activation functions.}
	\label{fig:doe_cylinder}	
\end{figure}
\begin{figure}[p]
	\centering
	\includegraphics[width=0.7\linewidth]{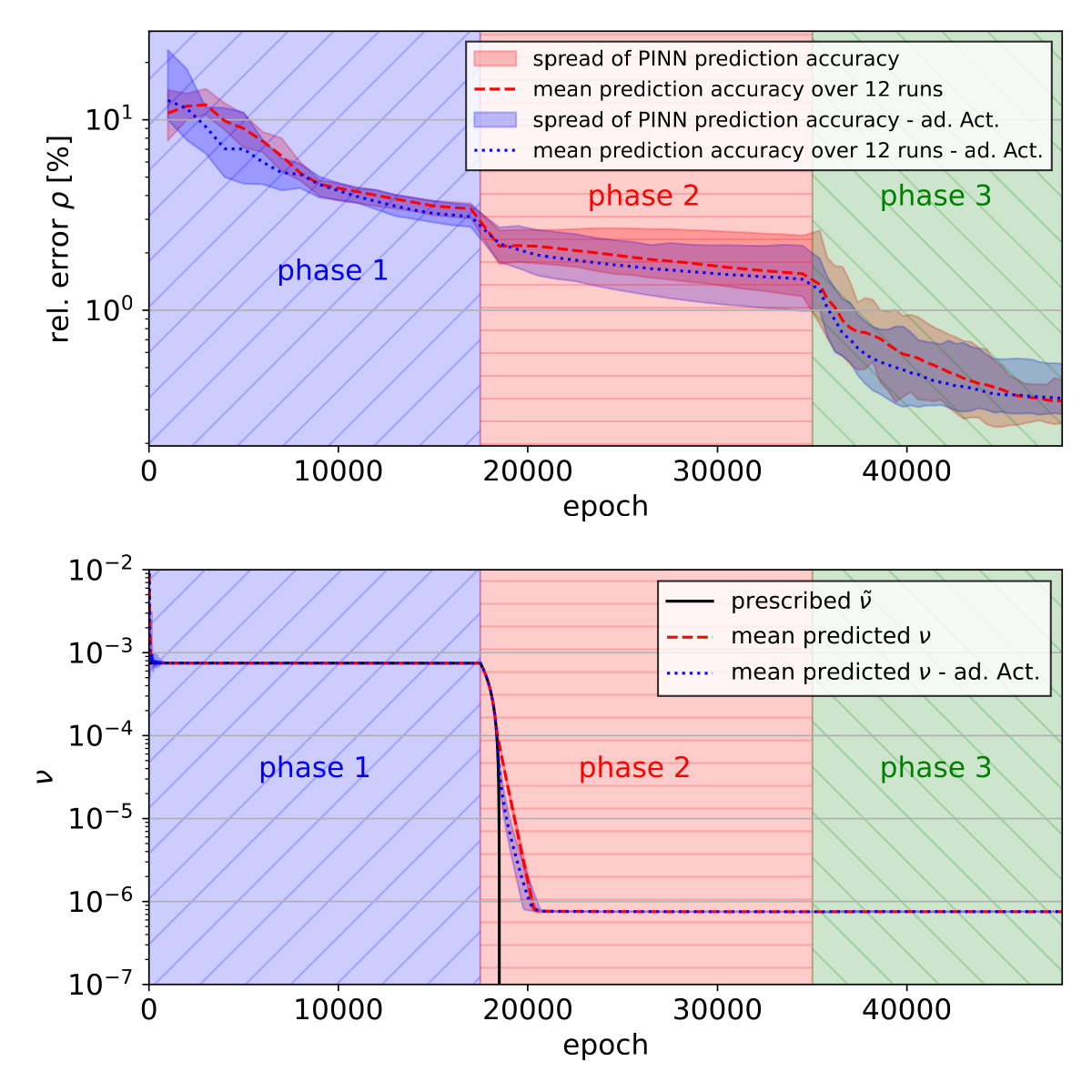}
	\caption{\comment{(updated) }Error, prescribed and predicted viscosity during training for flow around ellipse. Fig~(a) also shows the
	spread of prediction accuracy for 12 random initializations of the network.}
	\label{fig:cylinderTrainHist}
\end{figure}

For certain problems it can be observed that PINNs can perform inconsistently, depending on the random initialization
of network parameters at the start of the training. To ensure the consistency of our proposed method we analyze the 
accuracy over 12 training runs with different random initialization seeds. Fig.~\ref{fig:cylinderTrainHist}~(a) shows the
mean error in the density field during training. The best and worst prediction accuracy over the 12 runs 
(i.e. the spread of the predictions) is also indicated.
While differences occur between different initializations during early training phases, all models converge to a similar 
final accuracy. 
Fig.~\ref{fig:cylinderTrainHist}~(b) shows the prescribed viscosity factor $\tilde{\nu}$ as well as the 
predicted mean viscosity $\nu$ during training. Besides a spike in the first few epochs, the prescribed and predicted
viscosity are identical during the first phase. In phase 2 we see that the viscosity is comparatively quickly reduced.
After about 2500 epochs, it remains at a constant value around $\nu = 10^{-6}$.  
This indicates that this lower viscosity is sufficient for stabilizing the training and that $\nu$ can be reduced 
relatively fast during phase 2 without causing any instabilities. This is to be expected for this relatively stable 
subsonic problem.
However, as shown in \ref{sec:non-parametric-cylinder}, the initial viscosity in phase 1 is still required to converge 
to reasonable predictions which is even apparent for the non-parametric version of the problem. 
\added{On average we can only observe marginal improvements in accuracy and convergence speed, when using adaptive
	 activation functions instead of the fixed hyperbolic tangent activations.}
For a quantitative assessment of the errors, we consider the pressure coefficient $C_\mathrm{p}$, the local Mach number
$M$ and the density $\rho$ for 100 quasi-random parameter values in the parameter space.
For each parameter set, the mean absolute difference to the reference solution was calculated for a small square
that contains the cylinder ($-0.5<x<0.5; -0.5<y<0.5$). Fig.~\ref{fig:doe_cylinder} shows the resulting absolute errors
for the density. These errors are then normalized with the range of values of the reference solution for each 
individual quantity.
The final relative errors in Tab.~\ref{tab:errors} are the mean over all datasets and all 12 runs.
\added{The relative errors confirm that adaptive and non-adaptive activation functions perform very similarly 
	for this example.}
For a detailed explanation of the error calculation see Sec.~\ref{sec:apx_simulations_and_errors}. 
The hyperparameters are not optimized and higher accuracies may be possible when employing hyperparameter optimization 
e.g. on the network shape and learning rate.\par
For all presented results, the Python package SMARTy \cite{Bekemeyer.2022} was used, which is a toolbox for surrogate 
modeling and other data driven tasks. SMARTy supports Tensorflow and PyTorch as backends for the creation and training 
of neural network models. With the tensorflow backend the model is trained in 24 hours on a single 
NVIDIA~A100~graphics~card. 
All the presented models were using double precision for the floating point numbers. Once trained, the
evaluation of the model is fast. The prediction of the model at $300.000$ points takes about $0.5\;\mathrm{s}$.

\vspace{.5cm} 
\section{Parametric Oblique Shock}
\label{sec:parametric_oblique}
As a supersonic test case we consider the Oblique Shock problem with a variable inflow Mach number of $M\in(2, 3)$. 
The test case describes a scenario where a supersonic inflow is deflected by a wedge with deflection 
angle $\theta=10^\circ$. The resulting attached shock originates from the corner of the wedge. We define the shock 
angle with respect to the surface of the wedge as $\delta$. The shock angle is a unique function of the deflection 
angle and the incoming Mach number $M_1=M_\infty$ as stated by the $\theta$-$\beta$-$M$ relation. The field variables
after the shock can be calculated analytically from $M_1$ and $\theta$ \cite{Anderson.2011}.\par
This problem has been solved with PINNs in a non-parametric version with $M=2$ in \cite{Mao.2020} and \cite{Jagtap.2020}. 
For a comparison of the non-parametric results with and without adaptive viscosity, see 
\ref{sec:Apx_non_parametric_results}. To the best of our knowledge this is the only other forward supersonic test 
case which has been solved with PINNs for the two-dimensional compressible Euler equations.
PINNs have however been applied to other inverse supersonic problems \cite{Jagtap.2022} where shock
locations are already given by solution data that is provided inside of the physical domain.\par
Here, we solve the forward problem in the continuous parameter space for $M_\infty =(2, 3)$.
We use a total of $100000$ points for the evaluation of the residual and the viscosity penalty loss. 
$80000$ of these points are uniformly sampled in the three-dimensional input space 
$(x, y, M_\infty)\in \Omega \times (2, 3) = (0, 1)\times(0, 1)\times (2, 3)$. An additional 10.000 points are 
uniformly sampled on the upper $(x, y, M_\infty) \in \Omega \times \{3\}$ and lower bound 
$(x, y, M_\infty) \in \Omega \times \{2\}$ of the parameter space to enhance the accuracy towards the borders. 
The Halton sequence \cite{Halton.1960} is used for generating all three point sets. Note that, in contrast to previous 
works, we do not require clustered training points \cite{Mao.2020} or domain decomposition \cite{Jagtap.2020} for 
accurate predictions. This is advantageous, because no previous knowledge of the solution is required for the point 
generation or for the decomposition.
Dirichlet boundary conditions are applied at the top and left surface. No-flux wall boundary conditions are used 
for the bottom boundary. Since the shock originates from the bottom left corner, we increase the boundary point density
for $x\in(0, 0.1)$ and $y\in(0, 0.1)$ as sketched in Fig.~\ref{fig:BCs}. As before, an initial prescribed viscosity 
factor of $\tilde{\nu}=7.5\cdot 10^{-4}$ is used. The network consists of 7 layers with layers of 30 neurons and $\tanh$
activations.
\added{We compare fixed and adaptive hyperbolic tangent activation functions but freeze the trainable parameters
	of the activations in phase~3 as explained in Sec.~\ref{sec:parametric_cylinder}.}
The loss term weights are $\beta=1$ and $\gamma=5$. A detailed overview of training parameters is 
shown in \ref{tab:training_parameters_oblique}. Again, no additional data of the solution is incorporated into the loss, 
besides the boundary conditions and we are strictly solving the forward problem.\par
\begin{figure}[p]
	\centering
	\includegraphics[width=1\textwidth]{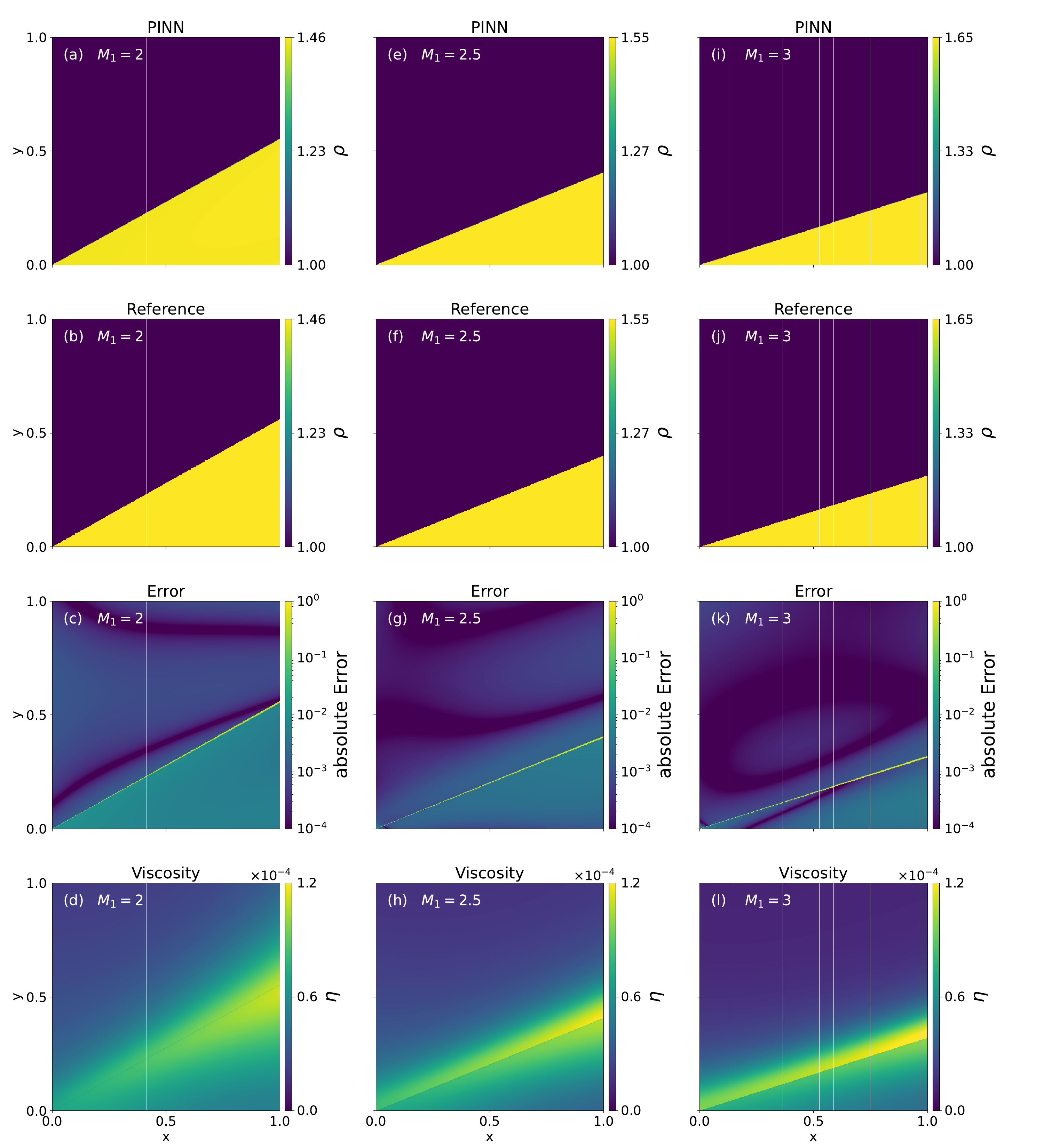}
	\caption{\comment{(updated) }Comparison of parametric PINN solution \added{using adaptive actiation functions} for oblique shock
			 test case with the analytical reference solution 
			 for different Mach numbers. The absolute errors between the reference and the PINN solution are 
			 shown in Figs.~(c),~(g) and (k). Figs.~(d),~(h)~and~(l), show the artificial viscosity $\eta$.}
	\label{fig:obliqueOverview}
\end{figure}
\begin{figure}
\centering
\includegraphics[width=0.7\linewidth]{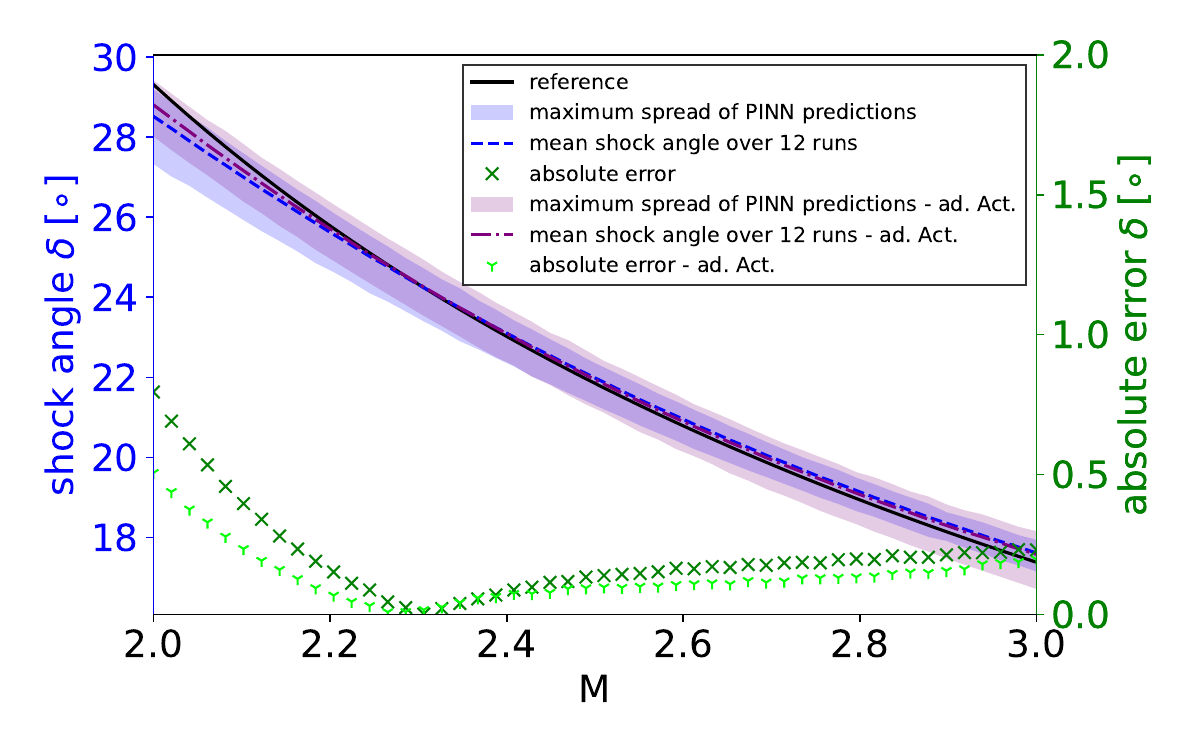}
\caption{\comment{(updated) }Mean and spread of predicted shock angles. The blue curves show the angle $\delta$ (for the definition see 
		 Fig.~\ref{fig:BCs}). The green curve shows the absolute error which is the absolute difference between the two 
	     blue curves.}
\label{fig:obliqueAngleErrors}
\end{figure}
\begin{figure}[p]
	\centering
	\includegraphics[width=0.75\linewidth]{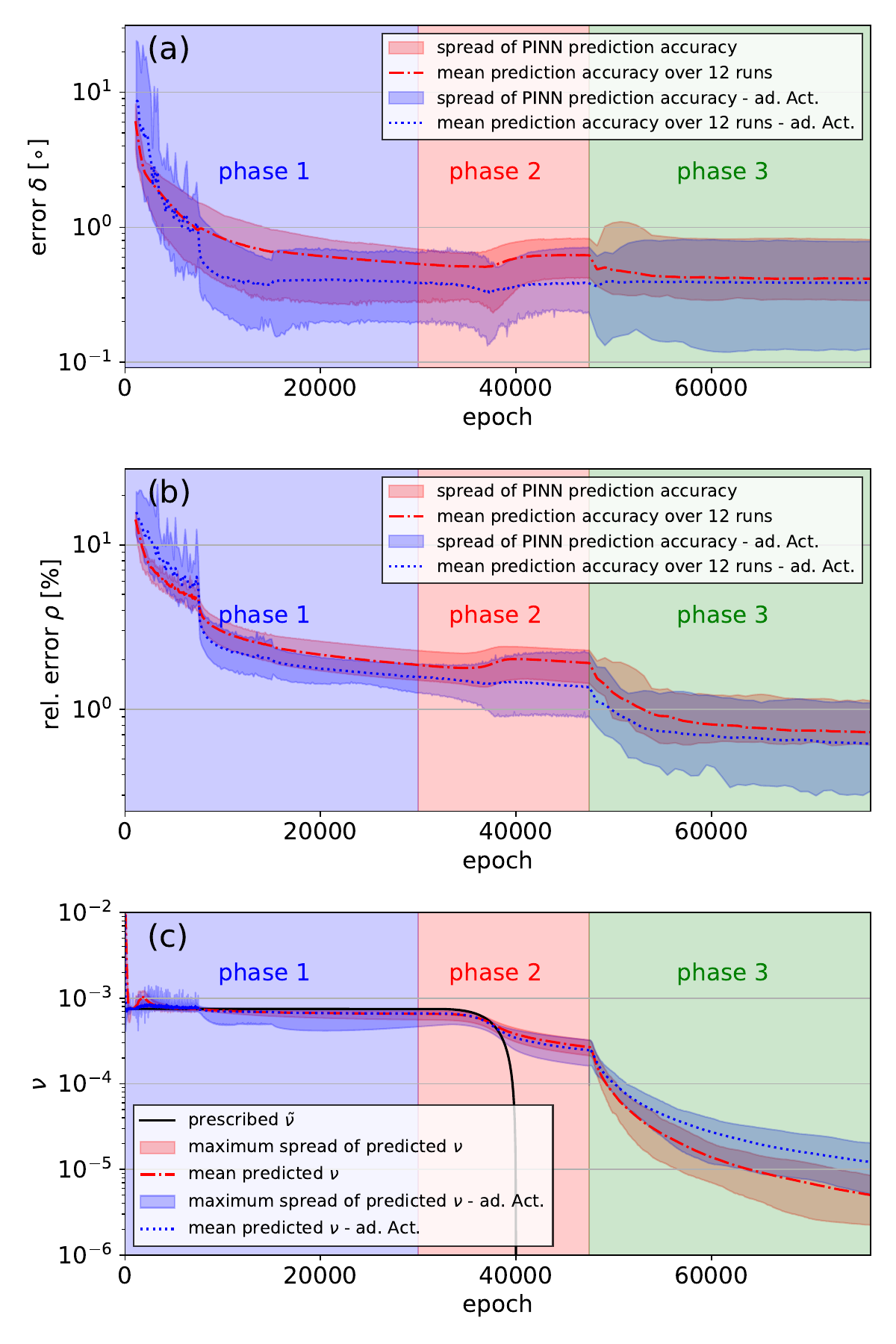}
	\caption{\comment{(updated) }Error history and artificial viscosity factor $\nu$ during training of PINN for parametric oblique shock 
			 problem. The three phases of the applied training procedure are highlighted.}
	\label{fig:obliqueTrainHistory}
\end{figure}
\begin{figure}
	\centering
	\includegraphics[width=1\linewidth]{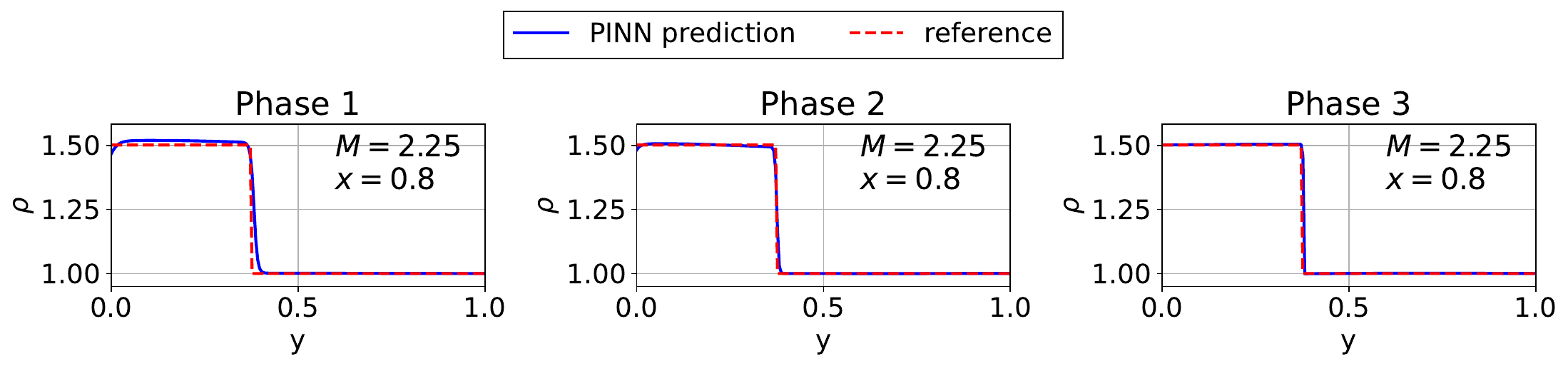}
	\caption{\comment{(updated) }Exemplary cross-section through density field at $M_\infty=2.25$ shows how shock becomes less 
		     dissipative after reducing the viscosity during the training. \added{Results were obtained using
	         the adaptive activation functions.}}
	\label{fig:shockduringtraining}
\end{figure}

Fig.~\ref{fig:obliqueOverview} provides an overview of the result for three different Mach numbers in comparison to the 
analytical solution. The field values before and after the shock are accurately predicted and the shock is well resolved. 
Slight inaccuracies in the shock angle are visible. The bottom row shows the artificial viscosity. Due to the local 
adaptivity, the dissipation is increased close to the shock. This shows that the proposed method is able to locally 
identify regions which require additional viscosity for stabilization. In this sense, the network is able to take over 
the role of so called pressure sensors, which are used in classical CFD methods to switch to first-order schemes near 
shock locations.
Since PINNs can perform inconsistently, depending on the random initialization of network parameters at 
the start of the training, we analyze the accuracy over 12 training runs with different random initialization seeds. 
The following plots highlight the mean prediction over those 12 runs and the largest and smallest values (i.e. the
spread of the predictions). The usual quantities like standard deviation and quantiles are not calculated due to 
the limited number of 12 training runs.
As an integral indicator of prediction accuracy we consider the shock angle $\delta$ (see Fig.~\ref{fig:BCs}).
Fig.~\ref{fig:obliqueAngleErrors} shows an overview of predicted shock angles for the entire parameter space. Overall 
the error of the shock angle is lower than one degree. The errors are larger at the lower bound of the parameter space 
towards $M_\infty=2$. The spread of the predictions over the 4 training runs indicates that the results are generally
within a one degree neighborhood of the analytical solution \added{both with and without adaptive 
activation functions. In this example, one can see that the adaptive outperform the fixed activation functions.}
Fig.~\ref{fig:obliqueTrainHistory}~(a) shows how the 
error for the angle delta changes during the training. We see that the error in the angle does not improve 
significantly after phase 1 and that the spread of predicted angles in fact increases. However,
Fig.~\ref{fig:shockduringtraining} shows that during the second and third training phase, the shock becomes much 
sharper and thus approximates the expected analytical result better. This is also confirmed by the decrease in the 
relative density errors (see Fig.~\ref{fig:obliqueTrainHistory}~(b)). Note that the final training phase 3 with LBFGS
is crucial to obtain good accuracy.
\added{The error history shows that on average, the adaptive activation functions can accelerate convergence in phase~1 and 2.
In phase 3 the most inaccurate runs with and without adaptive activations are very similar, while the most accurate runs
are improved for the adaptive activations.}
Fig.~\ref{fig:obliqueTrainHistory}~(c) shows the prescribed viscosity value $\tilde{\nu}$ and the mean (over the domain)
predicted viscosity $\nu$ during the training. Contrary to the subsonic test case we can clearly see the adaptivity of
viscosity and how it is only loosely coupled to the prescribed value. Early in the training we can see an adaptive
increase beyond the prescribed value which is accompanied by a fast convergence during the first few thousand epochs. 
Then, during phase 2, the predicted viscosity is again reduced less steeply than the prescribed value which indicates 
that more viscosity is necessary than prescribed, to stabilize the training during the reduction phase. During phase 3
the predicted viscosity decreases more steeply. The final values are between $\nu=2\cdot10^{-5}$ and $\nu=2\cdot10^{-6}$ and 
thus higher than for the subsonic test case. \added{The viscosity for the adaptive activations is on average
slightly increased.}
The necessity for more viscosity during the training is to be expected since we are dealing with 
a supersonic flow that involves a shock. Compared to the previous subsonic test case, this problem should be more 
unstable and require more dissipation.\par
For a quantitative comparison of relative errors with the subsonic problem, we again consider the pressure coefficient 
$C_\mathrm{p}$, the local Mach number $M$ and the density $\rho$ for 20 linearly distributed Mach number values with 
$M\in[2, 3]$.
For each parameter, the mean absolute difference to the reference solution was calculated for the entire domain. 
This difference was then normalized with the range of values of the reference solution for each individual quantity. 
The calculation of errors and uncertainties is described in more detail in Sec.~\ref{sec:apx_simulations_and_errors}. 
The hyperparameters are not optimized and higher accuracies may be possible when employing hyperparameter optimization 
e.g. on the network shape and learning rate. The use of adaptive activations generally leads to improved accuracies,
especially for the Mach number $M$. Compared to the subsonic problem \ref{sec:parametric_cylinder}, we see 
slightly increased errors within the same order of magnitude. The bounds of the error are higher, mainly due to
the fact that accuracies are worse close to the lower Mach number $M=2$, while the errors of the subsonic problem are 
consistently low for the entire parameter space (c.f. Sec.~\ref{sec:apx_simulations_and_errors}).\par
Again, the models are implemented with SMARTy \cite{Bekemeyer.2022} using the tensorflow backend and the total training 
time for one run on a NVIDIA~A100 graphics card is about 23 hours. The prediction at $300000$ points takes about 
$0.5\;\mathrm{s}$.
\begin{table}
	\centering
	\caption{\comment{(updated) }Comparison of relative errors for different field variables.}
	\label{tab:errors}
	\begin{tabular}{|l|cc|cc|}
		\hline
		\multirow{1.3}{*}{
			\textbf{Sec.}}
		&\multicolumn{2}{c|}{\makecell[c]{\textbf{\ref{sec:parametric_cylinder}. Parametric Flow} \\
			\textbf{around Ellipse}}} 
		&\multicolumn{2}{c|}{
		\makecell[c]{\textbf{\ref{sec:parametric_oblique}. Parametric Oblique}\\ 
			\textbf{Shock}}}
		\\ \hline
		 & $\sigma$ & $\sigma_{\mathrm{adapt}}$ & $\sigma$ & $\sigma_{\mathrm{adapt}}$\\
		$C_\mathrm{p}$ & $(0.23\pm0.06)\%$ & $(0.25\pm0.08)\%$ &  $(0.7\pm 0.7)\%$ & $(0.52\pm 0.61)\%$\\
		$M$ & $(1.0\pm0.1)\%$ & $(1.0\pm0.11)\%$ & $(1.3\pm 0.9)\%$ & $(0.7\pm 0.7)\%$\\
		$\rho$ & $(0.33\pm0.11)\%$ & $(0.34\pm0.13)\%$ & $(0.62\pm 0.62)\%$ & $(0.5\pm 0.6)\%$\\ \hline
	\end{tabular}
\end{table} 
\section{Conclusion}
\label{sec:conclusion}
To summarize, we propose a novel physics-informed neural network training procedure to approximate parametric
solutions to the stationary compressible Euler equations. Parameters are considered as additional input dimensions 
of the network. Furthermore, we add a dissipative term to the equations to stabilize the training process. 
Our proposed method locally predicts the necessary viscosity. An additional penalty loss term is used to control and 
reduce the viscosity during the training so that the resulting solution is non-dissipative. 
We obtain accurate results on a subsonic test case with a parametric Mach number and boundary shape as well as a 
supersonic test case with parametric Mach number. Adaptive activation functions perform similar on the subsonic
test case and outperform the fixed activation functions on the supersonic test case.
The proposed method is easy to implement and outperforms vanilla PINNs without viscosity 
(c.f.~\ref{sec:Apx_non_parametric_results}) which have so far rarely been applied successfully, for more than 
one-dimensional forward problems governed by the Euler equations without requiring previous solution knowledge (e.g. 
for data, clustered points or solution based domain decomposition). Thus, the presented method may open up new 
possibilities for the use of PINNs for similar inviscid problems, which were previously unsolvable, using vanilla PINNs. 
Compared to finite volume reference simulations, we achieve errors on the order of less than $2\%$ for the pressure, 
velocity and density field, for both test-cases while using no additional data besides boundary conditions.\par
\added{While the shown results are overall promising, one might consider the training cost to be a limiting 
	factor of the method. The viscous term introduces additional, second-order derivatives which increase the 
	computational effort per epoch due to additional automatic differentiation calls.
	In addition, the used neural networks have to be comparatively large
	because they need to capture the solution of four field variables and the artificial viscosity factor. 
	With regards to the additional training cost due to the viscous term, we consider this to be a sensible compromise
	because, as seen in $\ref{sec:Apx_non_parametric_results}$, convergence without viscous terms seems to be
	unreliable at best. This is to be expected due to the mathematical nature of hyperbolic conservation laws and we 
	know from classical numerical methods that additional stabilizing methods such as artificial viscosity are required 
	to obtain unique and physical solutions.
	In the future we plan to reduce the network size by decomposing the domain into subsets which are appoximated by
	smaller individual networks. Even though we obtain accurate solutions with a global network for the entire domain, 
	we expect that domain decomposition will be able to further increase the accuracy, enable faster parallel training 
	and to reduce the requirements for network size. Especially for larger domains, 
	as well as transient and higher-dimensional problems, domain decomposition approaches may be imperative because 
	very large global networks are limited in resolving local effects and training convergence becomes increasingly 
	difficult. Fortunately, since our methodology just requires the network to predict the local viscosity factor $\nu$ 
	and a simple modification of the loss function, it can easily be incorporated into existing domain decomposition 
	frameworks. Decomposition methods with explicit interface losses, such as extended PINNs~\cite{Jagtap.2020d}, would require additional continuity 
	conditions for the viscosity factor at the domain interfaces. Therefore, soft domain decomposition approaches such 
	as augmented PINNs~\cite{Hu.2023} or finite basis PINNs~\cite{Moseley.2023} might be favorable.
	Overall , we see various possibilities to further improve numerical efficiency of the described approach in 
	the future.	Moreover, we also have to take into account that inference times are on the order of $1\;\mathrm{s}$ 
	and that due to the parametric formulation, one is able to obtain solutions for different parameter conditions with 
	a single trained PINN model. This makes parametric PINNs viable in specific scenarios which require real-time 
	evaluation.}
The proposed implementation of artificial viscosity is simplistic and one can think of many ways to improve this 
approach. Similarly to artificial matrix-valued artificial viscosity schemes in classical CFD methods
\cite{Turkel.1989, Swanson.1987, Langer.2014b, Langer.2012}, a yet to be developed matrix-valued viscosity model for 
$\eta$ might reduce the influence of the dissipative term during training, while maintaining its stabilizing properties.
Also, an alternative regularization scheme based on an entropy criteria has been proposed for solving
hyperbolic problems with PINNs which might be an additional measure to avoid unphysical results, when solving the
compressible Euler equations, especially at higher Mach numbers \cite{Patel.2022, Jagtap.2022}. However, our initial 
tests indicate that this methodology is not sufficient to stabilize our analyzed forward problems on its own.\par
The presented approach of parametric boundary conditions can, in theory, be extended to more parameters (e.g. 
with variable Mach number, shape and angle of attack for airfoils). Therefore, it could be applicable for tasks such
as design optimization which traditionally require many solver evaluations. Whether a sufficient level of accuracy 
can be reached for more complex multi-parameter problems remains to be seen. An aspect of PINNs that has not been 
considered in this work is that higher fidelity solution data can be incorporated into the objective function as 
an additional loss term. This opens the possibility to directly combine PINNs with classical solvers or to even 
incorporate experimental data into the loss. For parametric problems, such a hybrid-data-driven approach may improve
accuracy and even speed up the convergence during training. This flexibility may open up new possibilities for 
physics-informed reduced order modeling for higher dimensional parametric flows.\par
An additional point of future interest is the behavior of the presented approach in the transsonic regime, where
velocities exceed the speed of sound only locally. 
\added{Furthermore, we want to consider transient problems. We expect that this requires additional measures 
	like decomposition of the temporal domain (see e.g.~\cite{Penwarden.2023}) while respecting temporal causality. 
	For transient problems we expect long training times because, when employing causality preserving temporal
	decomposition strategies, every temporal subdomain has to go through the training phases 1-3 successively, to 
	avoid an accumulation of viscous effects over time.}
\section*{Acknowledgements}
This work was supported by the Helmholtz Association's Initiative and Networking Fund on the HAICORE@FZJ partition.

\clearpage
\appendix
\section{Non-Parametric Problems}
\label{sec:Apx_non_parametric_results}
To highlight the necessity and the efficacy of artificial viscosity, we consider non-parametric versions of the 
previously analyzed problems and compare the results with and without adaptive artificial viscosity. 
\subsection{Flow around Cylinder}
\label{sec:non-parametric-cylinder}
\begin{figure}
	\centering
	\includegraphics[width=\linewidth]{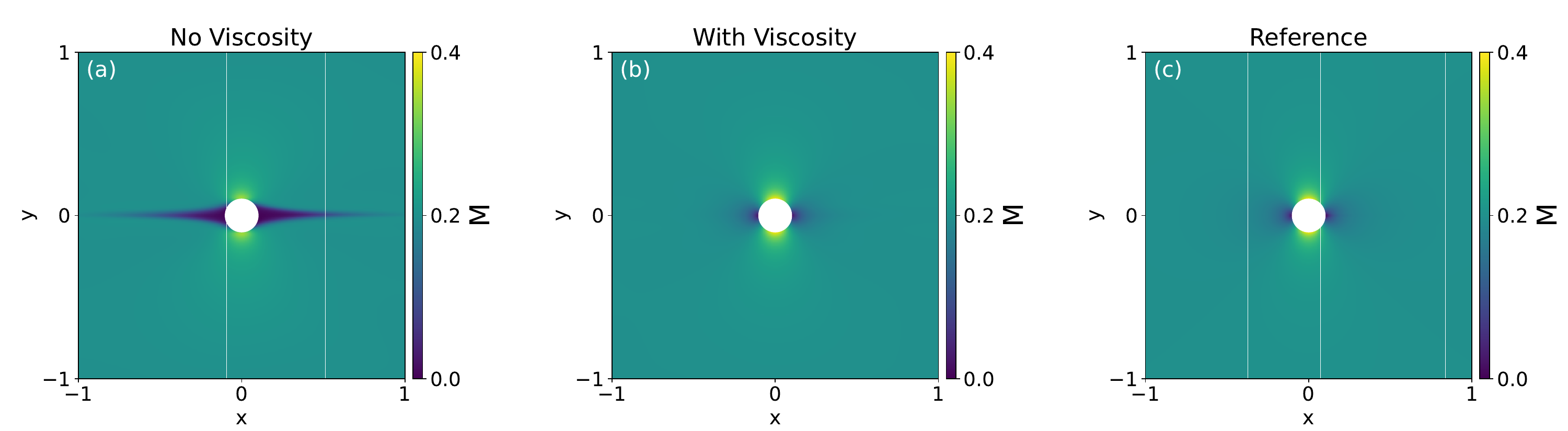}
	\caption{Comparison of PINN predictions for the local Mach number $M$ with and without artificial viscosity and the
		reference finite volume simulation.}
	\label{fig:apx_cylinder_compare_novisc}
\end{figure}

First, we investigate the flow around a cylinder at a constant Mach number of $M=0.2$. Again, the 
center of the cylinder is positioned at the origin inside the domain $\bm{x}= (x,y)\in \Omega = (-1,1)\times(-1,1)$ 
with a radius of $r=0.1$. Fig.~\ref{fig:apx_cylinder_compare_novisc}~(c) show a numerical reference solution of the
problem for a Mach number of $\mathrm{M}=0.2$. The solution was obtained using the CFD solver CODA~\cite{Kroll.2016}.
For additional information on the calculation of reference finite volume results, see 
\ref{sec:apx_simulations_and_errors}.\par
Similarly to before, no-flux boundary conditions are used for the cylinder wall and Dirichlet 
boundary conditions are enforced using Eq.~\eqref{eq:BoundaryLossEuler} for the external domain boundary.
A summary of the parameters within each training phase is shown in 
Tab.~\ref{tab:training_parameters_cylinder}. For the PINN without viscosity, we train for the same number of epochs 
with the ADAM and LBFGS optimizer as for the model with adaptive viscosity. The learning rate for ADAM is however set to
$lr=10^3$ for the entirety of the ADAM training. All other hyper parameters and the training points are identical\\
The training points for the residual loss are randomly sampled in the entire domain, using the quasi-random 
Halton sequence~\cite{Halton.1960}. For both point sets, the same number of points are used to represent the boundary 
of the physical domain and the cylinder ($N=N_\mathrm{\infty}=N_\mathrm{ob}$).\par
Figs.~\ref{fig:apx_cylinder_compare_novisc}~(a)-(b) show the resulting velocity fields given by the local Mach number.
The training parameters correspond to phase 1 in Tab.~\ref{tab:training_parameters_cylinder} .
The results without adaptive viscosity shows elongated, unphysical regions of low velocity before and after the cylinder.
In comparison, the model with adaptive viscosity agrees well with the reference simulation.\par
The training with SMARTy~\cite{Bekemeyer.2022} and the tensorflow backend on a single NVIDIA~A100 graphics cars takes about 
$6$~hours without viscosity and about $11$~hours for the model with viscosity, due to the increased effort for 
calculating the loss function.\par
\vspace{.5cm}
\subsection{Oblique Shock}
\label{sec:non-parametric-oblique}
We consider the oblique shock at a incoming Mach number of $M_\infty = 2$. The training points for the residual loss
are generated similarly to Sec.~\ref{sec:parametric_oblique} with the Halton sequence inside ob the physical domain
$(x,y)\in \Omega = (0,1)\times(0,1)$. Dirichlet boundary conditions are used for the left and upper boundary and 
no-flux wall conditions are used for the bottom boundary. No boundary condition is used on the right boundary.\par 
This exact forward problem has been solved with PINNs in~\cite{Mao.2020} and conservative PINNs in \cite{Jagtap.2020}. 
Hence, to demonstrate the efficacy of 
artificial viscosity, we try to reproduce the reported results of \cite{Mao.2020} by selecting similar hyper parameters 
for the neural network shape (7 layers with 20 neurons) and a similar number of uniformly distributed points 
(5000 for the residual loss) and compare the predictions with and without the adaptive viscosity for these exact hyper 
parameters. Note that we use uniformly distributed and no clustered training points because we do not want to assume 
any previous knowledge about the solution and the shock angle. We do however increase the boundary point density in the 
bottom left corner at the shock origin as schematically depicted in Fig.~\ref{fig:BCs}. Also the number of points on 
the boundaries overall increased to 2000, compared to 300 points in \cite{Mao.2020}. Again, both PINN models with and
without adaptive viscosity are trained for the same number of ADAM epochs (phase 1 and 2 combined for the viscous PINN) 
and LBFGS epochs.\par
\begin{figure}
	\centering
	\includegraphics[width=\linewidth]{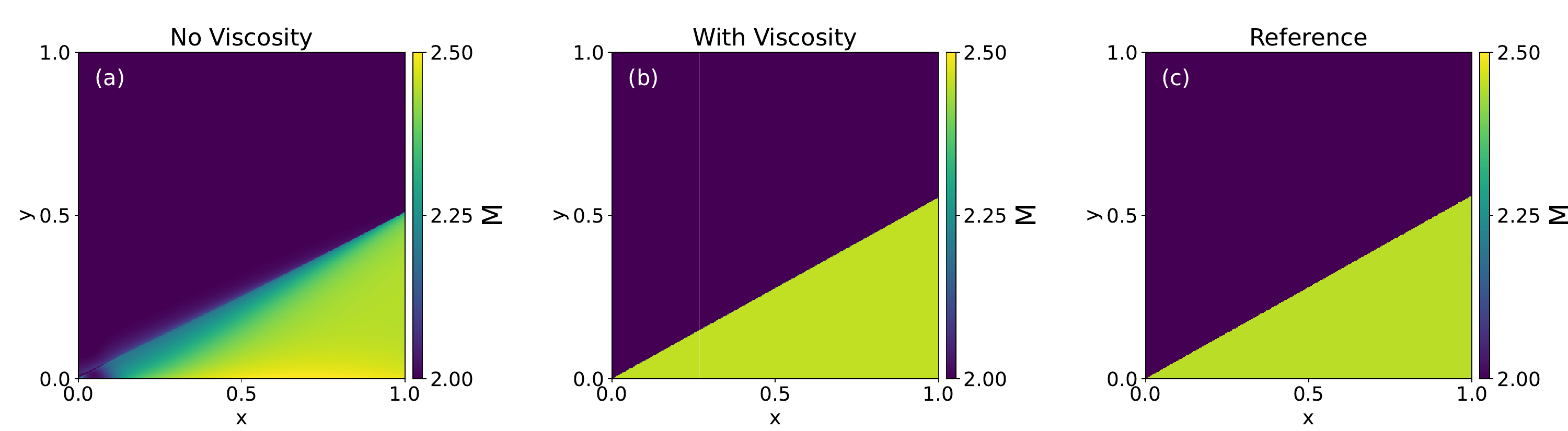}
	\caption{Comparison of results for the non-parametric oblique shock problem at $M_\infty = 2$.
		The local Mach number $M$ is shown for training runs with (Fig.~(b)) and without (Fig.~(a)) adaptive artificial
		viscosity in comparison to the reference finite volume result~(c). The hyper parameters used for (a) and (b) are
		the same. A search for better hyper parameters for (a) has not lead to significant improvements even when 
		employing dynamic loss term weighting~\cite{Maddu.2022, Wang.2021, Jin.2021} or adaptive activation 
		functions~\cite{Jagtap.2020b}.}
	\label{fig:apx_oblique_compare_novisc}
\end{figure}

Fig.~\ref{fig:apx_oblique_compare_novisc} shows the resulting density fields.\par
We are unable to obtain similar results for the inviscid PINN as shown in \cite{Mao.2020} even though the number of boundary 
points and the number of ADAM epochs have been increased. We have also searched for better hyper parameters and even 
employed adaptive activation functions and dynamic loss term weighting but were still unable to obtain qualitatively improved 
results. Nevertheless, even when compared to the shown results in \cite{Mao.2020}, we obtain highly improved results using
the proposed adaptive viscosity during training. The shock is much sharper and less dissipative, even without clustered training points. 

\clearpage
\section{Training Parameters}
\vspace{2cm}
\begin{center}
    \captionsetup{type=tabular}
    \captionof{table}{\comment{(updated) }Summary of PINN parameters for flow around ellipse.}
    \label{tab:training_parameters_cylinder}
    \centering
    \begin{tabular}{lcc}
        \hline
        \rule[-1ex]{0pt}{2.5ex}\textbf{Sec.} &
        \makecell[c]{\textbf{\ref{sec:non-parametric-cylinder}. Flow} \\
                     \textbf{around Cylinder}}  & 
        \makecell[c]{\textbf{\ref{sec:parametric_cylinder}. Parametric Flow}\\ 
                     \textbf{around Ellipse}} \\
        \hline
        \rule[-1ex]{0pt}{2.5ex} hidden layers & $8$ & $8$ \\
        \hline
        \rule[-1ex]{0pt}{2.5ex} neurons/layer & $20$ & $40$ \\
        \hline
        \rule[-1ex]{0pt}{2.5ex} activation  & (adapt.) $\tanh$ & (adapt.) $\tanh$ \\
        \hline
        \rule[-1ex]{0pt}{2.5ex} loss weights 
        & \makecell[ct]{$\beta = 1$\\$\gamma=5$} 
        & \makecell[ct]{$\beta = 1$\\$\gamma=5$}\\
		\hline
        \rule[-1ex]{0pt}{2.5ex} phase 1
        & \makecell[ct]{Adam\\ 
            $\tilde{\nu} = 7.5\cdot10^{-4}$\\
            $N=20000$\\
            $N_\mathrm{batch}= 2500$\\
            $10000$ epochs w.    $lr=10^{-3}$\\
            $20000$ epochs w.    $lr=10^{-4}$
        }								    
        & \makecell[ct]{Adam\\ 
            $\tilde{\nu} = 7.5\cdot10^{-4}$\\
            $N=100000$\\
            $N_\mathrm{batch}= 10000$\\
            $7500$ epochs w.    $lr=10^{-3}$\\
            $7500$ epochs w.    $lr=10^{-4}$\\
            $2500$ epochs w.    $lr=10^{-5}$}\\
        \hline
        \rule[-1ex]{0pt}{2.5ex} phase 2 & \makecell[c]{Adam\\ 
            reduce $\tilde{\nu} = 7.5\cdot10^{-4}$ to $\nu_0 = 0$\\
            $N=20000$\\
            $N_\mathrm{batch}= 2500$\\
        	$M_{\mathrm{red}}=2500$\\
            $k=4$\\
            $5000$ epochs w.   $lr=10^{-4}$\\
        	$5000$ epochs w.   $lr=10^{-5}$
        }					    						    
        & \makecell[c]{Adam\\ 
            reduce $\tilde{\nu} = 7.5\cdot10^{-4}$ to $\nu_0 = 0$\\
            $N=100000$\\
            $N_\mathrm{batch}= 10000$\\
            $M_{\mathrm{red}}=1000$\\
            $k=1$\\
            $17500$ epochs w.   $lr=10^{-5}$}\\	
        \hline
        \rule[-1ex]{0pt}{2.5ex} phase 3 & \makecell[c]{LBFGS\\ 
            $\tilde{\nu} = 0$\\
            $N=5000$\\
            $25000$ epochs}								    
        & \makecell[c]{LBFGS\\ 
            $\tilde{\nu} = 0$\\
            $N=30000$\\
            $30000$ epochs} \\
        \hline
    \end{tabular}
\end{center}
\clearpage

\newpage

\vspace{2cm}
\begin{center}
	\captionsetup{type=tabular}
	\captionof{table}{\comment{(updated) }Summary of PINN parameters for oblique shock problem.}
	\label{tab:training_parameters_oblique}
	\centering
	\begin{tabular}{lcc}
		\hline
		\rule[-1ex]{0pt}{2.5ex}\textbf{Sec.} &
		\makecell[c]{\textbf{\ref{sec:non-parametric-oblique}} \\
					 \textbf{Oblique Shock}}  & 
		\makecell[c]{\textbf{\ref{sec:parametric_oblique}. Parametric}\\ 
					 \textbf{Oblique Shock}}\\
		\hline
		\rule[-1ex]{0pt}{2.5ex} hidden layers & $7$ & $7$ \\
		\hline
		\rule[-1ex]{0pt}{2.5ex} neurons/layer & $20$ & $30$ \\
		\hline
		\rule[-1ex]{0pt}{2.5ex} activation  & (adapt.) $\tanh$ & (adapt.) $\tanh$ \\
		\hline
		\rule[-1ex]{0pt}{2.5ex} loss weights 
		& \makecell[ct]{$\beta = 1$\\$\gamma=5$} 
		& \makecell[ct]{$\beta = 1$\\$\gamma=5$}\\
		\hline
		\rule[-1ex]{0pt}{2.5ex} phase 1
		& \makecell[ct]{Adam\\ 
			$\tilde{\nu} = 7.5\cdot10^{-4}$\\
			$N=5000$\\
			$N_\mathrm{batch}= N$\\
			$12000$ epochs w.    $lr=10^{-3}$\\
			$18000$ epochs w.    $lr=10^{-4}$
		}								    
		& \makecell[ct]{Adam\\ 
			$\tilde{\nu} = 7.5\cdot10^{-4}$\\
			$N=100000$\\
			$N_\mathrm{batch}= 10000$\\
			$7500$ epochs w.    $lr=5\cdot10^{-4}$\\
			$7500$ epochs w.    $lr=10^{-4}$\\
			$15000$ epochs w.    $lr=2\cdot10^{-5}$}\\
		\hline
		\rule[-1ex]{0pt}{2.5ex} phase 2 & \makecell[c]{Adam\\ 
			reduce $\tilde{\nu} = 7.5\cdot10^{-4}$ to $\nu_0 = 0$\\
			$N=5000$\\
			$N_\mathrm{batch}= N$\\
			$M_{\mathrm{red}}=5000$\\
			$k=1$\\
			$20000$ epochs w. 	$lr=2\cdot10^{-5}$\\}						    
		& \makecell[c]{Adam\\ 
			reduce $\tilde{\nu} = 7.5\cdot10^{-4}$ to $\nu_0 = 0$\\
			$N=100000$\\
			$N_\mathrm{batch}= 10000$\\
			$M_{\mathrm{red}}= 10000$\\
			$k=4$\\
			$17500$ epochs w. 	$lr=2\cdot10^{-5}$}\\	
		\hline
		\rule[-1ex]{0pt}{2.5ex} phase 3 & \makecell[c]{LBFGS\\ 
			$\tilde{\nu} = 0$\\
			$N=5000$\\
			$10000$ epochs}								    
		& \makecell[c]{LBFGS\\ 
			$\tilde{\nu} = 0$\\
			$N=20000$\\
			$30000$ epochs} \\
		\hline
	\end{tabular}
\end{center}
\clearpage
\clearpage
\section{Details on Reference Simulations, Calculation of Errors and Shock Angles}
\label{sec:apx_simulations_and_errors}
\subsection{Parametric Flow around Ellipse}
All reference simulations for the ellipse were calculated using CODA \cite{Kroll.2016, Langer.2022}. 
CODA is the computational fluid dynamics (CFD) software
being developed as part of a collaboration between the French Aerospace Lab ONERA, the German Aerospace Center
(DLR), Airbus, and their European research partners. CODA is jointly owned by ONERA, DLR and Airbus. We use a
structured O-type grid with 200 surface nodes and 13200 cells. The residuals are converged to an order of $10^{-12}$.
For a quantitative evaluation of the accuracy of the parametric PINN in Sec.~\ref{sec:parametric_cylinder}, we compare
the solutions to 100 reference finite volume simulations in the parameter space 
$(M, a)\in (0.2, 0.4)\times(0.1, 0.2)$. The parameter sets sampled using the Halton sequence \cite{Halton.1960}.
Fig.~\ref{fig:doe_cylinder} shows the parameter values and the absolute errors for the density field. For each parameter
set, the errors are calculated for each field variable separately, by taking the mean of the absolute difference
over a small rectangle $(x, y)\in (-0.5, 0.5)\times(-0.5, 0.5)$ near the ellipse. The figure then depicts the mean 
error over all training runs. Overall the errors are all on the same order of magnitude and 
we see no significant outliers in the entire parameter space. The results are less accurate towards the bounds of 
the parameter space.\par
For the calculation of the relative errors in Tab.~\ref{tab:errors} and Fig.~\ref{fig:cylinderTrainHist} the previously 
calculated absolute errors are normalized with the range of the respective field variable, taken from the reference 
simulations. Then the mean over the parameter sets is taken. For Tab.~\ref{tab:errors} all 100 parameter sets and for 
Fig.~\ref{fig:cylinderTrainHist} a subset of 10 parameter sets is used. Finally, the mean over all 12 training runs is 
taken. The uncertainties in \ref{tab:errors} take the variance in accuracy at different locations in the domain, 
different parameter sets and the confidence with respect to the different simulation runs into account. 
\subsection{Parametric Oblique Shock}
For the oblique shock problem, the reference data is obtained from the analytical solution 
\cite[pp. 608-615]{Anderson.2011}). The relative errors in Fig.~\ref{fig:obliqueTrainHistory} and 
Tab.~\ref{tab:errors} are calculated similarly to the ellipse problem but taking the entire domain 
$(x, y)\in (0, 1)\times(0, 1)$ into account. A total number of 20 and 5 uniformly distributed $M_\infty$ values are 
considered for Tab.~\ref{tab:errors} and  Fig.~\ref{fig:obliqueTrainHistory}, respectively. Again, the uncertainties 
in Tab.~\ref{tab:errors} take the variance in accuracy at different locations in the domain, different parameter sets and 
the confidence with respect to the different simulation runs into account. The predicted shock angles in
Fig.~\ref{fig:obliqueAngleErrors} and Fig.~\ref{fig:obliqueTrainHistory} are determined with a root finding algorithm 
which looks for the mean reference value of the density before and after the shock ($(\rho_1+\rho_2)/2$). This value 
is searched in the PINN prediction field on a vertical line at $x=0.95$. This method has worked sufficiently fast and 
accurate for our analysis with errors $\ll 1^\circ$. The shock location is very consistent between the different field 
variables. Therefore, it is sufficient to only calculate the angle based on the density prediction. 
\clearpage
\bibliographystyle{elsarticle-num} 
\bibliography{Literature.bib}
\end{document}